\documentclass[12pt, ams,  onecolumn, amssymb,  floatfix, longbibliography]{revtex4-2}

\usepackage[utf8]{inputenc}
\usepackage{amsmath}
\usepackage{amsfonts}
\usepackage{amssymb}
\usepackage{graphicx}
\usepackage{bm}
\usepackage{float}
\usepackage{ulem}
\usepackage{cancel} 
\usepackage{slashed}
\usepackage{comment}
\usepackage{mathtools}

\usepackage{ulem}

\usepackage{enumitem}

\usepackage{subcaption}

\usepackage{fixltx2e,amsmath}
\usepackage[titletoc,toc]{appendix} 
\usepackage{extarrows}

\usepackage{natbib}
 
\newcommand{\mycomment}[1]{%
}%

\newcommand{\Tr}{\text{Tr}}

 \newcommand{\gX}{g_X}

 \newcommand{\wx}{w_x}

\newcommand{\pd}{\partial}

\newcommand{\nn}{\nonumber}

\newcommand{\PiM}{\hat{\Pi}  }

\newcommand{\CAtd}{ \tilde{C}_A   }

\usepackage[colorlinks=true,
            linkcolor=OrangeRed,
            urlcolor=blue,
            citecolor=gray]{hyperref}
 \hypersetup{citecolor=RoyalBlue} 
\usepackage[usenames,dvipsnames]{color}
\definecolor{dkgray}{RGB}{145,145,145}
\definecolor{violet}{RGB}{50,0,200}

\numberwithin{equation}{section}

\begin{document}

\title{Universal conductivity at a 2d superconductor-insulator transition: the effects of quenched disorder and Coulomb interaction}
\author{Chao-Jung Lee}
\affiliation{Department of Physics, California Institute of Technology, Pasadena, CA 91125, USA}
\author{Michael Mulligan}
\affiliation{Department of Physics and Astronomy, University of California, Riverside, CA 92521, USA}

\date{\today}

 \bigskip
 \bigskip
 \bigskip

\begin{abstract}
We calculate the zero-temperature universal electrical conductivity at a superconductor-insulator transition in two spatial dimensions.
We focus on transitions in the universality class of the dirty 3d XY model.
We use a dual model consisting of a single Dirac fermion at zero density coupled to a Chern-Simons gauge field in the presence of a quenched random mass, with or without an unscreened Coulomb interaction.
Our calculation is performed in a $1/N_f$ expansion, where $N_f$ is the number of Dirac fermions.
At zeroth order, the model exhibits particle-vortex self-dual electrical transport with $\sigma_{xx} \lesssim (2e)^2/h$ and small, but finite $\sigma_{xy}$. 
Corrections of ${\cal O}(1/N_f)$ due to fluctuations in the Chern-Simons gauge field and disorder produce violations of self-duality.
We find these violations to be milder when the Coulomb interaction is present.
\end{abstract}

\maketitle

\bigskip

\newpage

\setcounter{page}{1}

\definecolor{napiergreen}{rgb}{0.16, 0.5, 0.0}
\definecolor{officegreen}{rgb}{0.0, 0.5, 0.0}

\definecolor{seagreen}{rgb}{0.18, 0.55, 0.34}
\definecolor{sacramentostategreen}{rgb}{0.0, 0.34, 0.25}
\definecolor{upforestgreen}{rgb}{0.0, 0.27, 0.13}

\definecolor{tropicalrainforest}{rgb}{0.0, 0.46, 0.37}

\definecolor{viridian}{rgb}{0.25, 0.51, 0.43}

\definecolor{pakistangreen}{rgb}{0.0, 0.35, 0.0}

\definecolor{GreenForTableofCont}{rgb}{0.0, 0.5, 0.32}

\definecolor{navyblue}{rgb}{0.0, 0.0, 0.5}
\definecolor{persianblue}{rgb}{0.11, 0.22, 0.73}

	\definecolor{brickred}{rgb}{0.8, 0.25, 0.33}
 	\definecolor{brown(web)}{rgb}{0.65, 0.16, 0.16}

{  \hypersetup{linkcolor=persianblue}
   \tableofcontents
}

\newpage


\section{Introduction}
\label{introsection}

Continuous quantum phase transitions \cite{sachdev_2011,RevModPhys.69.315} are one of the most intriguing phenomena in condensed matter physics, due to the possibility of emergent behaviors that are different from the proximate phases.
Superconductor to insulator transitions (SITs) in disordered two-dimensional thin films, the focus of this paper, provide some of the best examples (\cite{Goldmanreview2010}, for review).
Here, as a tuning parameter, such as the charge density, disorder, or external magnetic field, is varied, the electrons in the disordered film transition between a superconducting and insulating ground state.
At a critical value of the tuning parameter, the electrons form a metal with finite, nonzero dc resistance (measured as the temperature $T \rightarrow 0$) on the order of the quantum of Cooper-pair resistance $R_Q = h/4e^2 \approx 6.45\ k\Omega/\square$, consistent with the theoretical prediction \cite{PhysRevB.40.546, PhysRevLett.64.587, PhysRevLett.65.923} that the critical resistivity is a universal amplitude.

This prediction is based on the hypothesis \cite{PhysRevLett.64.587, PhysRevLett.65.923} that such transitions are due to the phase disordering of the superconducting order parameter, rather than the loss of superconducting pairing amplitude \cite{PhysRevLett.86.1869}.
The critical properties of such a transition should then be described by a model of interacting charge-2e Cooper-pair bosons moving in a random potential in two spatial dimensions \cite{PhysRevLett.64.587, PhysRevLett.65.923}.
One of the more intriguing possibilities suggested by this model is that the transition might be self-dual, i.e., the critical Hamiltonian for Cooper-pair bosons on the brink of localization is the same as the dual critical Hamiltonian for vortices \cite{PESKIN1978122,PhysRevLett.47.1556,PhysRevB.39.2756} on the brink of condensation.
A consequence of self-duality is the so-called semicircle law for the dc electrical conductivity:
\begin{eqnarray}
\label{pvsymmetryrelation}
 \sigma^2_{xx}   +  \sigma^2_{xy} 
= \big(\frac{4e^2}{h}\big)^2 .
\end{eqnarray}
This relation is particularly interesting because it relates the dissipative ($\sigma_{xx}$) and nondissipative ($\sigma_{xy}$) parts of the conductivity to a universal constant, $R_Q$.
There is strong experimental evidence \cite{Breznay2016} that field-tuned transitions are self-dual with $\sigma_{xx} = 1/R_Q$ and $\sigma_{xy} = 0$.
This may be surprising, given the presence of the nonzero magnetic field, and suggests an emergent particle-hole symmetry \cite{FISHER1991553}.
Measurements \cite{Goldmanreview2010} of the longitudinal resistance near $R_Q$ at charge density or disorder tuned SITs are suggestive of an underlying self-duality, so long as the impurities are nonmagnetic.

The argument for self-duality within the boson model is indirect \cite{PhysRevLett.64.587, PhysRevLett.65.923}.
Here, we will consider an alternative description, given in terms of 3d quantum electrodynamics with a single Dirac fermion coupled to a Chern-Simons gauge field \cite{ChenFisherWu1993, BarkeshliMcGreevy2012continuous, PhysRevB.93.205116, PhysRevB.95.045118}.
In close analogy to theories \cite{jain_2007, Fradkinbook, PhysRevX.5.031027} for the half-filled Landau level, we will refer to the Dirac fermions in this description as composite fermions. 
In this composite fermion model, self-duality is a consequence of an unbroken particle-vortex symmetry that (loosely speaking) acts on the Hamiltonian of the model as a time-reversal symmetry and therefore fixes the composite fermion Hall conductivity to be zero \cite{PhysRevB.95.045118}.
(The composite fermion conductivity is distinct from the electrical conductivity. We will recall the correspondence between the two in \S \ref{electricalresponsesection}.)
This means that, so long as particle-vortex symmetry is preserved, the composite fermion theory will yield self-dual response.
The precise manner in which self-duality \eqref{pvsymmetryrelation} is realized depends on the specific value of the composite fermion longitudinal conductivity.
For instance, a composite fermion longitudinal conductivity of ${1 \over 2} R^{-1}_Q$ reproduces \eqref{pvsymmetryrelation} with $\sigma_{xy} = 0$.

Calculations of the longitudinal conductivity are notoriously challenging (e.g., \cite{PhysRevB.44.6883, PhysRevB.49.12115, damlesachdev97}). 
To gain insight, we consider a limit of the composite fermion theory that corresponds to lattice bosons at integer filling with charge-conserving disorder \cite{PhysRevB.40.546}.
In this limit, the critical properties are those of the dirty 3d XY model \cite{PhysRevB.40.546}, recently clarified in \cite{PhysRevB.101.144506}.
In the fermion dual, the Dirac composite fermions lie at zero density in the presence of a quenched random mass.
In \cite{PhysRevResearch.2.023303}, we showed that this theory admits renormalization group (RG) fixed points at finite disorder, with or without the Coulomb interaction, when the mean of the random mass is tuned to zero.
(Other symmetry classes of disorder have not, as yet, yielded accessible fixed points in this theory.)
These fixed points are controlled within a $1/N_f$ expansion, where $N_f$ is the number of fermion flavors.
Related works studying the effects of quenched randomness on theories of Dirac fermions coupled to a fluctuating boson include \cite{PhysRevLett.80.5409, PhysRevB.73.155104, PhysRevB.77.195413, PhysRevB.95.235145, PhysRevB.95.235146, PhysRevB.98.195142}.

An unscreened Coulomb interaction is believed to be relevant to the experimentally-realized SITs \cite{PhysRevLett.64.587, PhysRevLett.65.923};
within the 3d XY model, the Coulomb interaction corresponds to a relevant perturbation.
General constraints on the transport properties of a self-dual SIT, with Coulomb interactions, were derived in \cite{PhysRevB.96.075127,PhysRevB.100.235150}.
The stability of these results to fluctuations in the Coulomb interaction, in the presence of disorder, is not addressed in these papers.
Here, we determine the electrical conductivity at the random-mass fixed points of \cite{PhysRevResearch.2.023303}, with or without the unscreened Coulomb interaction, to ${\cal O}(1/N_f)$, with our results given in Eqs.~\eqref{nocoulombresult} and \eqref{withcoulombresult} and Figs.~\ref{fig:sub1} and \ref{fig:sub2}.
At ${\cal O}(1/N_f^0)$, the theory exhibits self-dual electrical transport with $\sigma_{xx} \lesssim 1/R_Q$ and finite, nonzero $\sigma_{xy}$.
The leading ${\cal O}(1/N_f)$ corrections, due to fluctuations in the Chern-Simons gauge field and disorder, violate self-duality.
These violations are found to be smaller at the fixed point with unscreened Coulomb interaction.

There are two important things to note about our results.
First, the random mass does not preserve particle-vortex symmetry. 
This is not the underlying reason why we find nonzero composite fermion Hall conductivity: The ${\cal O}(1/N_f)$ violation of self-duality is due to the fluctuations of the Chern-Simons gauge field; there is a nonzero composite fermion Hall conductivity in the pure limit without disorder.
We are unaware of a calculable model of a disordered fixed point that is self-dual. 
Second, we calculate the conductivity in the phase-coherent regime $\hbar \omega/k_B T \rightarrow \infty$ (where $\omega$  is the measuring frequency).
This regime gives a universal conductivity tensor that, in general, differs from conductivity in the incoherent regime $\hbar \omega/k_B T \rightarrow 0$ \cite{damlesachdev97}.

The structure of this paper is as follows.
In \S \ref{set-up-our-models}, we define the model that we study.
In \S \ref{sec-DCF-conductiity}, we sketch the calculation of the composite fermion conductivity for general values of the coupling constants.
In \S \ref{sec-RG-fixed-pt}, we evaluate these composite fermion conductivities at two RG fixed points, corresponding to disordered quantum critical points with or without the unscreened Coulomb interaction, and use the duality dictionary to translate these quantum critical composite fermion conductivities to electrical conductivities.
In \S \ref{discussion}, we discuss our results.
Three appendices contain technical details of the calculations summarized in the main parts of the paper: Appendix \ref{appendix-gauge-propagator} contains the derivation of the effective gauge field propagator; Appendices \ref{appendix-random-mass-correction} and \ref{appendix-gauge-correction} contain details for the evaluation of loop diagrams.

\section{The Model}  \label{set-up-our-models}

In this section, we define the Dirac composite theory of the SIT.
See \cite{PhysRevResearch.2.023303} for additional details.

\subsection{The effective action}

The total Euclidean effective action for the Dirac composite fermion theory has three parts:  
\begin{eqnarray}
\label{totalaction}
S_{\text{tot} }  = S_0 +  S_c  + S_{\text{dis}}.
\end{eqnarray}
We will define each part in the following three paragraphs.

To begin, $S_0$ is a Euclidean action of quantum electrodynamics in 3d.
It consists of a two-component Dirac fermion $\psi$ that is coupled to a dynamical Chern-Simons gauge field $a_\mu$ ($\hbar = 2 e = 1$):
\begin{align}
  \label{S0-action}
S_0 & = \sum_{I=1}^{N_f}
\int d^2xd\tau \, 
\bar{\psi}_I \gamma^0  \,  \, (\pd_\tau - i \frac{g}{\sqrt{N_f}} \, a_\tau) \psi_I 
+
\bar{\psi}_I \gamma^j  \,   (  \pd_j - i \frac{g}{\sqrt{N_f}} \, a_j) \psi_I   \nn \\
& + \frac{  i\, \kappa }{2} a  da   
+
\frac{ i }{4\pi} (-2Ada + AdA).
\end{align}
In \eqref{S0-action}, $\bar{\psi} \equiv \psi^\dagger \gamma^0 $ and the $2 \times 2$ gamma matrices are chosen such that the anti-commutator $\{ \gamma^\mu, \gamma^\nu   \}= 2 \delta^{\mu \nu} \bm{1}_{2\times 2}$, where $\mu, \nu  \in  \{ 0,1,2 \}  =  \{ \tau,x,y \}$. 
We follow the Einstein summation convention, with, for instance, the spatial indices $j \in \{x,y\}$ summed over above.
(Being in Euclidean signature, upper and lower indices are equivalent and will be used interchangeably.)
The Chern-Simons term $ada \equiv \epsilon^{\mu \nu \lambda} a_\mu \pd_\nu a_\lambda$ (and similarly for the other Chern-Simons terms), where the Levi-Civita symbol $\epsilon^{012} \equiv +1$. 
Here, $N_f$ is the number of Dirac fermion flavors.
$S_0$ is dual to the critical 3d XY model when $N_f = 1$ and the Chern-Simons coefficient $\kappa= \frac{1}{4\pi}$ \cite{ChenFisherWu1993, BarkeshliMcGreevy2012continuous, PhysRevB.95.045118}, with the Dirac mass playing the role as the critical tuning parameter.
We take $N_f > 1$ in order to study the theory within a controlled $1/N_f$ expansion.
$A_\mu$ is a probe, nondynamical electromagnetic gauge potential that serves to define the electrical current and corresponding electrical conductivity of the model.
The gauge coupling $g$ is set to unity in the infrared; we keep $g$ general for the moment. 

Next, we introduce the Coulomb interaction. 
Following \cite{PhysRevB.92.235105, PhysRevResearch.2.023303}, the Coulomb interaction dualizes into the composite fermion theory as
\begin{eqnarray}
S_c
=
\frac{1}{2} 
\int \frac{d^3k}{(2\pi)^3} a_T (k) 
\Big(\,
|\bm{k}| \, \wx
\Big) a_T(-k)   ,
\end{eqnarray}
where we introduced the effective Coulomb coupling  $w_x \equiv \frac{e^2_* }{8\pi^2}$. 
Here, $e_\ast$ is the charge of the Cooper-pair bosons (nominally, $e_\ast = 2 e$).
The Fourier space integration measure is $d^3k \equiv dk_x dk_y dk_0$, where $k_0$ is the zero-temperature Matsubara frequency. 
We adopt the following notation throughout this paper:
\begin{eqnarray}
\bm{k}=(k_x,k_y), \; k_\mu =(k_0,k_x,k_y) , \;
|\bm{k}| = \sqrt{k_x^2+k_y^2} , \;
k =\sqrt{ k_0^2 + k_x^2+k_y^2}.
\end{eqnarray}
The ``$T$" subscript indicates that $a_T$ is the transverse component of the Chern-Simons gauge field: $a_T$ is related to the Cartesian components $(a_x, a_y)$ via $a_x(q) =i   \frac{q_y}{|\bm{q}|} a_T(q)$ and $a_y(q) = -i   \frac{q_x}{|\bm{q}|} a_T(q)$, provided we choose Coulomb gauge, wherein the longitudinal component of $a_j$ is set to zero.

Finally, we turn to the introduction of the quenched randomness. 
Previous analyses \cite{PhysRevResearch.2.023303,PhysRevB.95.235146} show that if all types of disorder (quenched random couplings that couple to composite fermion bilinears), regardless of the symmetry they individually preserve, are added to the theory, the system flows to strong coupling, out of reach of any analytic control. 
However, if charge-conjugation symmetry is imposed, only a random mass term $m(\bm{x})\bar{\psi} \psi$ is allowed and the theory flows to an accessible disordered fixed point \cite{PhysRevResearch.2.023303}. 
Applying the standard replica trick to disorder average, the action picks up the term,
\begin{eqnarray}
S_{\text{dis}} = 
\frac{-1}{2}
\int d^2x d\tau d\tau' \sum_{r,r' =1}^{ n_r }
g_m \, (\bar{\psi}_r \psi_r)_{x,\tau}  \,
(\bar{\psi}_{r'} \psi_{r'} )_{x,\tau'}   \,  ,
\end{eqnarray}
where $r,r'$ are the replica indices and the number of replica $n_r$ is to be set to zero in the last step of any calculation. 
The replica indices are not important in the later calculations, so we will not write them out explicitly anymore (this is also why we didn't include them in the earlier parts of the total action). 
The parameter $g_m$ is the disorder strength, which is always non-negative.
The random mass disorder does not preserve particle-vortex symmetry.
This will be apparent later when we discuss the quantum critical conductivity of the model.

\subsection{The effective gauge field propagator}

In the Cartesian basis, the one-loop gauge field self-energy induced by the Dirac fermions is
\begin{eqnarray}
\Pi_{\mu \nu}^{1-loop}    \label{one-loop-Pi}
=\frac{ -g^2 }{16}  
\frac{ \delta_{\mu \nu}  k^2 - k_\mu k_\nu }{k}.
\end{eqnarray}
This self-energy is ${\cal O}(N_f^0)$, i.e., the same order as the bare terms in the total action that are quadratic in the gauge field.
Adding this self-energy to $S_c$ and the Chern-Simons term in $S_0$ involving $a_\mu$ only, we obtain the one-loop gauge field action,
\begin{eqnarray}
&&   \label{Sgauge-propagator}
S_{\text{gauge}} 
=
\frac{1}{2} \int \frac{d^3k}{(2\pi)^3}
\begin{pmatrix}
a_0 &  a_T
\end{pmatrix}_{k}
\begin{pmatrix}
\frac{ \gX^2   \bm{k}^2  }{ k   } &   i \kappa \, |\bm{k}|  \\
  i \kappa \, |\bm{k}|  & \,  \gX^2  \, k  + |\bm{k}| \,\wx  \\
\end{pmatrix}
\begin{pmatrix}
a_0 \\  a_T
\end{pmatrix} _{-k},
\end{eqnarray}
where $ \gX^2 \equiv \frac{g^2}{16}$.
This action defines the one-loop corrected gauge field propagator. 

The computation we summarize in the next section is performed with the gauge field, expressed in the Cartesian basis.
We therefore translate the gauge field propagator, defined by Eq.~\eqref{Sgauge-propagator}, to the Cartesian basis:
\begin{eqnarray}
&&    \label{D00}
D_{00}(k_0,\bm{k})
=
\frac{1}{ \bm{k}^2  }
\frac{1}{\kappa^2+ \gX^4  
  + \frac{\wx  \gX^2 \,|\bm{k}|}{  k  }   }
( \gX^2  \, k  + |\bm{k}| \, \wx  )  \\
&&   \label{D0i-anti-NoCoul} 
D_{0i}(k_0,\bm{k}) = -D_{i0}  
=
\frac{1}{ \bm{k}^2  }
\frac{1}{\kappa^2+ \gX^4  
  + \frac{ \wx   \gX^2 \,|\bm{k}|}{ k }   }
\,
(  \frac{  \kappa \,     \epsilon_{ij}k_j  }{1} )   \\
&&   \label{Dij}
D_{ij}(k_0,\bm{k})
=
\frac{1}{\kappa^2+ \gX^4  
  + \frac{ \wx    \gX^2 \,|\bm{k}|}{  k   }   }
(\frac{ \gX^2      }{  k   })  
(\delta_{ij}- \frac{k_i k_j }{\bm{k^2}})
\end{eqnarray}
To obtain $D_{\mu \nu}$ above, we have used the gauge fixing term $\frac{(\nabla a)^2}{2\lambda}$.
See Appendix \ref{appendix-gauge-propagator} for details.
 
\section{Composite fermion conductivity} 
\label{sec-DCF-conductiity}

In this section, we sketch the computation of the composite fermion conductivity, to two-loop order. 
The details of this calculation are in Appendices \ref{appendix-random-mass-correction} and \ref{appendix-gauge-correction}.
This composite fermion conductivity is related to the electrical conductivity in the next section.

\subsection{Definition of the composite fermion conductivity}

The composite fermion conductivity $\sigma^\psi_{ij}(\omega, {\bf p})$ is determined by the real-frequency retarded gauge field self-energy $\Pi^R_{ij}$, which is related to the imaginary-time self-energy $\Pi^{\rm tot}_{ij}$ as follows:
\begin{eqnarray}
\sigma^\psi_{ij}(\omega, \bm{p} )  =
\frac{i}{\omega}
\Pi^R_{ij}(\omega+ i 0^+, \bm{p} ) 
=
\frac{i}{\omega}
\Pi^{\rm tot}_{ij}( i p_0 \to \omega+ i 0^+, \bm{p} ).
\label{sigma-ij-conductivity}
\end{eqnarray}
Our definition of $\sigma_{ij}^\psi$ includes the contributions from the composite fermions; it does not include the tree-level Chern-Simons term.
We decompose $\Pi^{\rm tot}_{ij}$ into symmetric $\Pi^{\rm tot}_{(\text{S})}$ and anti-symmetric $\Pi^{\rm tot}_{(\text{A})}$ components as
\begin{eqnarray}
&&
\Pi^{\rm tot}_{\mu \nu} (p)
=
(\delta_{\mu \nu } - \frac{p_\mu p_\nu}{p^2})  \;    \Pi^{\rm tot}_{(\text{S})}  (p)
+
\frac{  p_\kappa   }{ |p|  } \,  \epsilon_{ \kappa \mu \nu}    
 \Pi^{\rm tot}_{(\text{A})} (p).   
 \label{Pi-munu-parameterize}
\end{eqnarray}

To extract the symmetric component of $\Pi^{\rm tot}_{ij}$, we will set $(\mu, \nu) = (x,x)$ and evaluate:
\begin{eqnarray}
\Pi^{\rm tot}_{(\text{S})}  = \frac{1}{ 1- \frac{p_x^2}{p^2}} \Pi^{\rm tot}_{xx} .
\end{eqnarray}
The anti-symmetric component will be obtained by contracting $\Pi^{\rm tot}_{ij}$ with $\epsilon_{\mu \nu \lambda} p^\lambda$:
\begin{eqnarray}
\Pi^{\rm tot}_{(\text{A}) }= \frac{p_\lambda  \epsilon_{\mu \nu \lambda}}{ 2 |p|} 
 \Pi^{\rm tot}_{\mu \nu}. 
 \label{ASym-Projection}
\end{eqnarray}

\subsection{Two-loop corrections to the conductivity}

\begin{figure}[h!]
  \centering
    \includegraphics[scale=0.5]{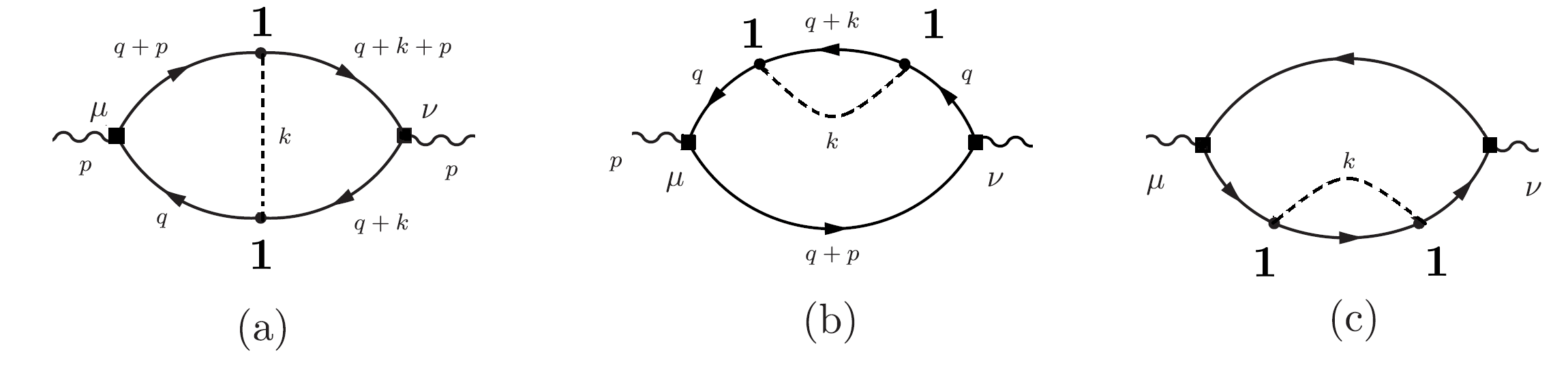}   
\caption{Two-loop diagrams with random mass corrections, denoted as $\PiM_{ij}$. } 
  \label{Figure-abc-gm}
\end{figure}

\begin{figure}[h!]
  \centering
    \includegraphics[scale=0.5]{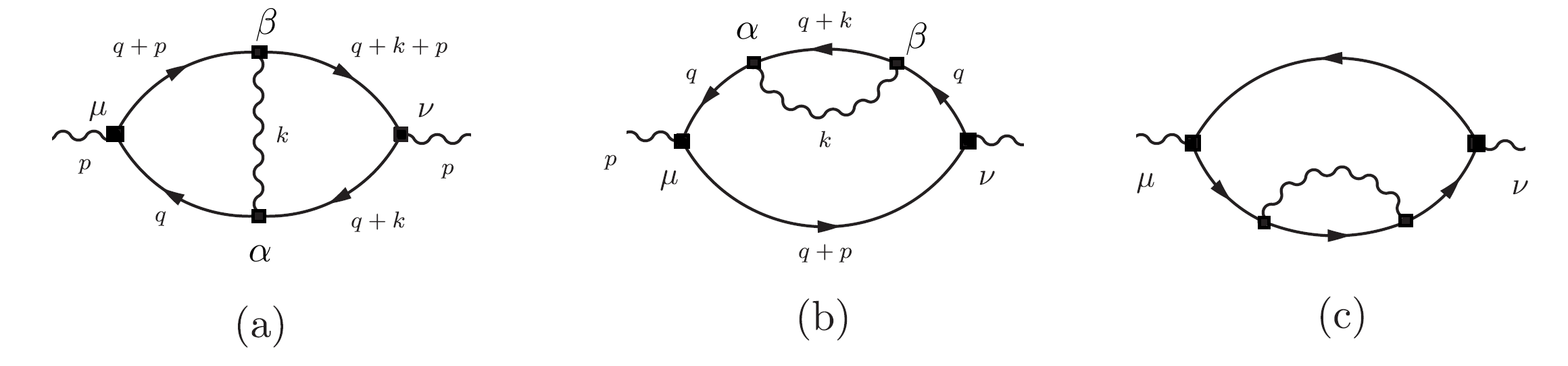}   
\caption{Two-loop diagrams with gauge field corrections, denoted as $\Pi_{ij}$. }
  \label{Figure-abc-gauge}
\end{figure}

We now calculate the composite fermion conductivity to two-loops, obtained from the two-loop corrections to $\Pi^{\rm tot}_{ij}$ by Eq.~\eqref{sigma-ij-conductivity}.
The one-loop corrections have already been included in the 1-loop corrected gauge boson self-energy.
The two-loop corrections result from self-energy corrections due to the random mass $\PiM_{ij}$ to ${\cal O}(g_m)$ and the fluctuations of the gauge boson $\Pi_{ij}$ to ${\cal O}(1/N_f)$.
The diagrams contributing to these corrections are shown in Figs.~\ref{Figure-abc-gm} and \ref{Figure-abc-gauge}. 
In each figure, diagrams (b) and (c) are equal.

We start with the corrections due to the random mass.
The diagrams in Fig.~\ref{Figure-abc-gm} are: 
\begin{eqnarray}
&&\PiM^{(a)}_{\mu \nu} (p) = - (\frac{ig}{\sqrt{N_f}} )^{2}\,N_f 
\int_{k, q}
\Tr[ \gamma_\mu G(q) G(q+k) \gamma_\nu G(q+k+p) G(p+q)] \times 2\pi  g_m \delta(k_0),  \label{Pi-a-mass} \quad \quad \quad  \\
&& \PiM^{(b+c)}_{\mu \nu} (p) = -2 (\frac{ig}{\sqrt{N_f}} )^{2} \,N_f
\int_{k,q}
\Tr[ \gamma_\mu G(q) G(k+q) G(q) \gamma_\nu  G(p+q)   ] 
\times 2\pi  g_m \delta(k_0)   ,
 \label{Pi-bc-mass} 
\end{eqnarray}
where we have used the shorthand $\int_{k,q} \equiv \int \frac{d^3k}{(2\pi)^3}
\int \frac{d^3q}{(2\pi)^3}$ and the trace is performed over the Dirac indices.
Here, $G(k)$ is the composite fermion propagator:
\begin{align}
-i G(k) = {k_0 \gamma^0 + k_i \gamma^i \over k_0^2 + k_i k^i}.
\end{align} 
To evaluate the gauge field diagrams in Fig.~\ref{Figure-abc-gauge}, we replace the disorder-induced four-point interaction by the gauge field propagator and disorder strength coupling by the fermion-gauge field coupling:
\begin{eqnarray}
&&   \label{Pi-a-gauge}
\Pi^{(a)}_{\mu \nu } (p)
=  - \frac{ g^4}{N_f}  
\int_{k, q}
 \Tr[ \gamma_\mu G(q) \gamma_\alpha G(q+k) \gamma_\nu G(q+k+p) \gamma_\beta G(p+q)]
\, D_{\alpha \beta}(k)    ,   \qquad   \\
&&   \label{Pi-bc-gauge}
\Pi^{(b+c)}_{\mu \nu} (p)
= -\frac{2 g^4}{N_f}  
\int_{k, q}
\Tr[ \gamma_\mu G(q) \gamma_\alpha  G(k+q) \gamma_\beta G(q) \gamma_\nu  G(p+q)   ] \,
D_{\alpha \beta } (k)    ,   \qquad
\end{eqnarray}
where $D_{\alpha \beta }$ is the gauge field propagator.

If the Coulomb interaction is present, the 3d Euclidean symmetry that rotates the temporal and spatial directions into one another is lost in the gauge field propagators; this symmetry is already broken by the random mass.
(This is due to the fact that the Coulomb interaction, considered as an instantaneous interaction, is fundamentally due to an electron density-density interaction that is mediated by the temporal component of the electromagnetic gauge field.
We have translated this interaction into the composite fermion theory \cite{PhysRevResearch.2.023303}.)
This makes the evaluation of these integrals by a naive Feynman parameterization difficult.
Instead, we re-express all terms in the integrands involving $q$ using partial fractions in a way that allows us to carry out the UV-finite integrals over $q$, after which we perform the $k$ integrals.
The lengthy calculations that do this are relegated to Appendices \ref{appendix-random-mass-correction} and \ref{appendix-gauge-correction}.
We now summarize the results.

The only nonzero contribution of the random mass diagrams is to $\hat \Pi_{xx}$:
\begin{eqnarray}
&&
\PiM^{(a+b+c)}_{xx} (p_0, \bm{p}=0)
=
3 \times 
(-1) (\frac{ig}{\sqrt{N_f}} )^{2}\,N_f\, g_m
\times \frac{1}{2\pi}
\times  \frac{|p_0|}{96}  .
\end{eqnarray}
This result was first obtained by Thomson and Sachdev in \cite{PhysRevB.95.235146}.

The gauge field diagrams result in nonzero contributions to both the symmetric and anti-symmetric components of $\Pi_{ij}$.
The anti-symmetric component is
\begin{eqnarray}                                \label{Pi-abc-gauge-Asym}
&& \Pi^{(a+b+c)}_{(\text{A}) }(p)    =
 - \frac{ 1 }{  2  |p|}
  \frac{ g^4}{N_f}     
  \int_{-1}^{1} dz
\, 
p^2 \,
\frac{ -z + (-1 + z^2) \text{ArcTanh}[z]    }{   8 \pi^2 \,z} 
 \times
 \frac{1}{ A_X + B_X \sqrt{1-z^2} }    , \qquad      \\
 && A_X \equiv  \frac{ \kappa^2+ \gX^4 }{ \kappa    }   , \;\;
 B_X \equiv   \frac{ \wx \gX^2  }{\kappa}   .
\end{eqnarray}
The symmetric component is
\begin{align}                              \label{Pi-abc-gauge-Sym}
\Pi^{(a+b+c)}_{(\text{S}) }(p_0, \bm{p}=0)   
& =
- \frac{g^4}{N_f}
 \, |p|  
\int_{-1}^{1} dz
\frac{f^{\wx}_0(z) + f^{\wx}_0(-z)  }{2  }    ,     \\
    \label{fwx-avg}
\frac{f^{\wx}_0(z) + f^{\wx}_0(-z)  }{2}      
& \equiv 
\frac{
-5 \,C_Y \,  z (-1 + z^2) + z \sqrt{1-z^2} (9 + z^2)} { 192 \pi^2 \, z \, [\; A_Y\sqrt{1-z^2} + B_Y(1-z^2)    \;\;  ]}   \cr 
& + 
   \frac{6 (-1 + z^2) (C_Y - C_Y\,  z^2 + 2 \sqrt{1-z^2}) \, \text{arctanh}[z]}
{ 192 \pi^2 \, z \, [\; A_Y\sqrt{1-z^2} + B_Y(1-z^2)    \;\;  ]} ,
\end{align}
where
\begin{eqnarray}
A_Y \equiv  \frac{ \kappa^2+ \gX^4  }{ \gX^2  } , \qquad
B_Y \equiv  \wx , \qquad
C_Y \equiv  \frac{ \wx  }{ \gX^2}.
\end{eqnarray}
As a consistency check, we note the precise agreement of our results in Eqs.~\eqref{Pi-abc-gauge-Asym} and \eqref{Pi-abc-gauge-Sym} for the gauge field diagrams, evaluated at $w_x = 0$, with the earlier computations of these same diagrams, computed in the absence of the Coulomb interaction, in \cite{SPIRIDONOV1991109, Huh:2015aa,Giombi:2016fct}:  
\begin{eqnarray}
&&  
\Pi^{(a+b+c)}_{(\text{A}) }(p; \wx =0) 
  =   \frac{ 1 }{ 2 |p|}
  \frac{g^4}{N_f}      
   \frac{ \kappa   }{\kappa^2+ \gX^4     }
 \;( p^2  \; \frac{4+ \pi^2}{32\pi^2}  
  \;  )  ,
  \\
&&
\Pi^{(a+b+c)}_{(\text{S}) }(p ; \wx =0, \kappa =0)
=   \frac{ - g^4}{N_f} \times  0.00893191.
\end{eqnarray}

Plugging these results into the definition Eq.~\eqref{sigma-ij-conductivity} using $\Pi^{\rm tot}_{ij} = \PiM_{ij} + \Pi_{ij}$, we find that to $\mathcal{O}(\frac{1}{N_f},g_m)$ the composite fermion conductivity equals
\begin{eqnarray}
&&  \label{sigma-CF-xx}
\sigma^{\psi}_{xx} (\omega) = \frac{1}{16} -  
   \frac{3g_m}{96\pi} 
  +  \frac{i }{\omega}  
      \Pi^{(a+b+c)}_{(\text{S}) }(|p_0| \to i \omega, \bm{p}=0)  ,
\\
&&  \label{sigma-CF-xy}
\sigma^{\psi}_{xy}(\omega) = 
 \frac{i }{  \omega   } 
 \Pi^{(a+b+c)}_{(\text{A}) }( |p_0| \to i \omega , \bm{p}=0  )   .
\end{eqnarray}

\section{Quantum critical electrical transport}  
\label{sec-RG-fixed-pt}

\subsection{RG flow and critical composite fermion conductivity}
	
\begin{figure}[h!]
  \centering
    \includegraphics[scale=0.5]{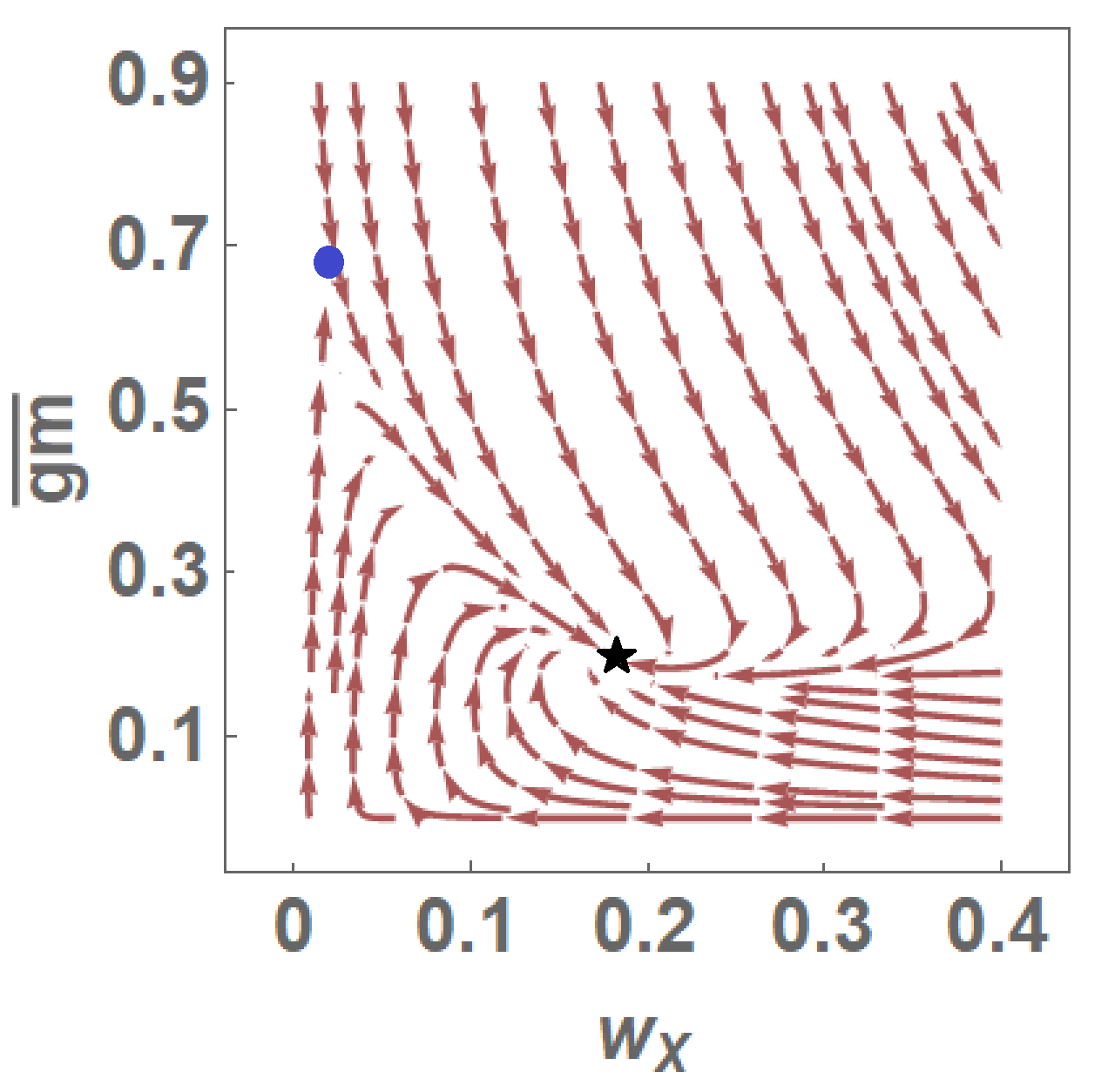}   
\caption{RG flow of the strength of the random mass ($g_m = 2 \pi \overline{g_m}$) and Coulomb coupling ($w_x)$ evaluated at $N_f =2$. 
The circle denotes the fixed point without the Coulomb interaction.
The star is the fixed point with finite Coulomb interaction. }
  \label{RG-flow-Betawx-Beta-gm}
\end{figure}

The perturbative RG flows for this system $S_{\rm tot}$ \eqref{totalaction} were found in \cite{PhysRevResearch.2.023303}. 
In that study, we allowed for the possibility of a dissipative Coulomb interaction.
We will not consider this possibility here.
With the Fermi velocity set equal to unity, the beta functions ($\beta_X \equiv - \mu \frac{\pd X}{\pd \mu} $) are
\begin{eqnarray}
&&   \label{beta-wx}
 \beta_{w_x} = w_x (z-1)  = w_x  [ \frac{g_m}{2\pi} -F_w(w_x,\kappa)    \;]  ,  \\
&&    \label{beta-gm}
\beta_{g_m}
=2 g_m (z-1 - \gamma_{\bar{\psi} \psi  }) 
= - 2 g_m  [  \frac{g_m}{2\pi}  + F_m(w_x, \kappa)   \;]  .
\end{eqnarray}
The dynamical exponent $z$, fermion anomalous dimension $ \gamma_{\bar{\psi} \psi  }$, and loop-functions $F_w, \, F_m$ are defined as follows:
\begin{align}
 z & = 1 + \frac{g_m}{2\pi} - F_w(w_x, \kappa)  ,   \\
\gamma_{\bar{\psi} \psi  } & = 2\frac{g_m}{2\pi} +F_m -F_w       , \\
F_w ( \wx , \kappa)
& =
\frac{1}{4\pi^2 N_f } \int_{-\infty}^{\infty} dy
\frac{ g_1(2y^2-1)   + \wx   \sqrt{1+y^2}  }
{(1+y^2)^2 [ \sqrt{1+y^2} (g_1^2+ \kappa^2)  + g_1 \wx    \;   ]}      , \\
F_m  ( \wx , \kappa) & = \frac{1}{4\pi^2 N_f} 
\int_{-\infty}^{\infty} dy
\Big[
\frac{ g_1(-2y^2-3)   -   \wx   \sqrt{1+y^2}  }
{(1+y^2)^2 [ \sqrt{1+y^2} (g_1^2+ \kappa^2)  + g_1 \wx   \;   ]}  \cr
& +
\frac{   \sqrt{1+y^2} (g_1^2 - \kappa^2)  + g_1 \wx    \;    }
{ 2(1+y^2)  [ \sqrt{1+y^2} (g_1^2+ \kappa^2)  + g_1 \wx    \;   ]^2    }  
\Big],
\end{align}
where $g_1 \equiv  g^2_X = \frac{g^2}{16} $.
The dynamical critical exponent and fermion anomalous dimension determine the inverse correction length exponent: $\nu^{-1} = z - \gamma_{\bar{\psi} \psi  }$.
Fig.~\ref{RG-flow-Betawx-Beta-gm} depicts the RG flow diagram for these beta functions \eqref{beta-wx} and \eqref{beta-gm}.

As Fig.~\ref{RG-flow-Betawx-Beta-gm} indicates, a finite disorder fixed point exists whether or not the Coulomb interaction is present.
In the absence of the Coulomb interaction ($w_x = 0$), $\beta_{w_x}$ in Eq.~\eqref{beta-wx} vanishes.
At $w_x = 0$, the $y$ integral in Eq.~\eqref{beta-gm} can be evaluated analytically.
The solution $g_m^\ast$ is given by 
\begin{eqnarray}
\frac{g_m^*}{2\pi}
\equiv
-F_m(0,\kappa)
=
\frac{ g_1^2 (-3+ 16 g_1)+ (3+ 16 g_1)\kappa^2 }{12\pi^2 \,N_f  (g_1^2+ \kappa^2)^2}
\approx
\frac{ 1.41091  }{N_f},
\end{eqnarray}
where we plugged in $\kappa = \frac{1}{4\pi}$ in the right-most equality. 
Substituting these fixed point values of $w_x$ and $g_m$ into the expressions for the conductivity, Eqs.~\eqref{sigma-CF-xx} and \eqref{sigma-CF-xy}, we obtain at $\omega = 0$:
\begin{eqnarray}
&&
\sigma^{\psi}_{xx}    \label{sigma-xx-wx=0}
=
\frac{1}{16} - \frac{0.084774}{N_f}     \qquad   (\wx=0)  ,  
 \\
&&
\sigma^{\psi}_{xy}   \label{sigma-xy-wx=0}
=
 \frac{ -0.170657   }{N_f}             \qquad   (\wx=0).
\end{eqnarray}
The correlation length and dynamical critical exponents at this fixed point are \cite{PhysRevResearch.2.023303}
$( \nu , z) = (1, \frac{1.4}{N_f})$.
Eq.~\eqref{sigma-xx-wx=0} makes clear that we cannot reliably extrapolate $N_f \rightarrow 1$, since this would violate the general requirement that the longitudinal conductivity is non-negative.

In the presence of the Coulomb interaction ($w_x \neq 0$), we have to solve the flow equations in \eqref{beta-wx} and \eqref{beta-gm} together.
The fixed point solution is found to lie at
\begin{eqnarray}
&&   
(w_x^*  , g_m^*)  
\approx  
(0.184193 ,  2\pi \, \frac{ 0.393787  }{N_f})   .
\end{eqnarray}  
Using Eqs.~\eqref{sigma-CF-xx} and \eqref{sigma-CF-xy}, this gives the dc composite fermion conductivity:
\begin{eqnarray}
&& \sigma^{\psi}_{xx}   \label{sigma-xx-wx-Not=0}
=
\frac{1}{16}
+
\frac{ 0.0177677  }{ N_f }                     \qquad   (\wx \neq 0)   ,  
 \\
&& \sigma^{\psi}_{xy}       \label{sigma-xy-wx-Not=0}
=
\frac{- 0.0903845}{ N_f }                        \qquad   (\wx \neq 0),
\end{eqnarray}
The critical exponents at this fixed point are \cite{PhysRevResearch.2.023303}
$( \nu, z) = (1,1)  $.  

We observe that disorder, without the Coulomb interaction, suppresses the composite fermion longitudinal conductivity \eqref{sigma-xx-wx=0}.
Inclusion of the Coulomb interaction results in an enhancement of the composite fermion longitudinal conductivity \eqref{sigma-xx-wx-Not=0}. 
Disorder and the fluctuations of the Chern-Simons gauge field produce a nonzero composite fermion Hall conductivity, \eqref{sigma-xy-wx=0} and \eqref{sigma-xx-wx-Not=0}.
Note that the composite fermion Hall conductivity remains nonzero in the pure limit at leading order in $1/N_f$.

\subsection{Electrical response}
\label{electricalresponsesection}

To determine the quantum critical electrical transport at the fixed points described in the previous section, we must first recall the dictionary that relates the composite fermion and electrical conductivities \cite{PhysRevB.95.045118}.
The two-loop calculation described previously can be understood to give rise to the following quadratic effective action, in which the composite fermion is integrated out:
\begin{eqnarray}
S^{\text{eff}}_{\text{DCF}}   \label{Seff-a+A}
=
\int \frac{d^2k d\omega}{(2\pi)^3}
\frac{ -i \omega }{2}
\Big(
a_i  \,  \sigma^\psi_{ij} \,  a_j
+
{1 \over 2 \pi} \epsilon_{t ij} (\frac{1}{2} a_i a_j -2A_i a_j +A_i A_j  )
\Big).
\end{eqnarray}
Here, we have used our gauge freedom to set $a_t = A_t = 0$.
Terms that are higher-order in the gauge field $a_i$ are ignored.
We integrate over $a_i$ to obtain the effective electrical response action:
\begin{eqnarray}
S^{\text{eff}}_{\text{DCF}}    \label{Seff-A}
=
\int \frac{d^2k d\omega}{(2\pi)^3}
\frac{ -i \omega }{2}
\epsilon_{t ij}  A_i \,    \sigma_{ij}      \, A_j ,
\end{eqnarray}
where the electrical conductivity is
 \begin{eqnarray}
&&   \label{conduct-xx-relation}
\sigma_{xx} = 
 \frac{ \sigma^{\psi}_{xx} }
  { (\sigma^{\psi}_{xx})^2 + (\sigma^{\psi}_{xy}+\frac{1}{4 \pi})^2} \cdot \big({1 \over 2\pi}\big)^2  ,    \\
&&     \label{conduct-xy-relation}
 \sigma_{xy} = 
 {1 \over 2 \pi} - 
 \frac{ \sigma^{\psi}_{xy}+\frac{1}{4 \pi}   }
  { (\sigma^{\psi}_{xx})^2 + (\sigma^{\psi}_{xy}+\frac{1}{4 \pi})^2}  \cdot \big( {1 \over 2 \pi} \big)^2  .
\end{eqnarray}
An immediate consequence of this dictionary is that
\begin{eqnarray}
\sigma_{xx}^2 + \sigma_{xy}^2
=
\frac{  (\sigma^\psi_{xx})^2 + (\sigma^\psi_{xy}- \frac{1}{4 \pi})^2    }
{ (\sigma^\psi_{xx})^2 + (\sigma^\psi_{xy}+ \frac{1}{4 \pi})^2 } \cdot \big({1 \over 2 \pi} \big)^2.
\end{eqnarray}
This shows that self-dual electrical transport \eqref{pvsymmetryrelation} occurs when the composite fermion Hall conductivity $\sigma_{xy}^\psi = 0$.
Our system, however, is not self-dual, since $\sigma_{xy}^\psi \neq 0$ (see \eqref{sigma-xy-wx=0} and \eqref{sigma-xy-wx-Not=0}).

\begin{figure}
\centering 
\begin{subfigure}{.3\textwidth}
  \centering
  \includegraphics[scale=0.4]{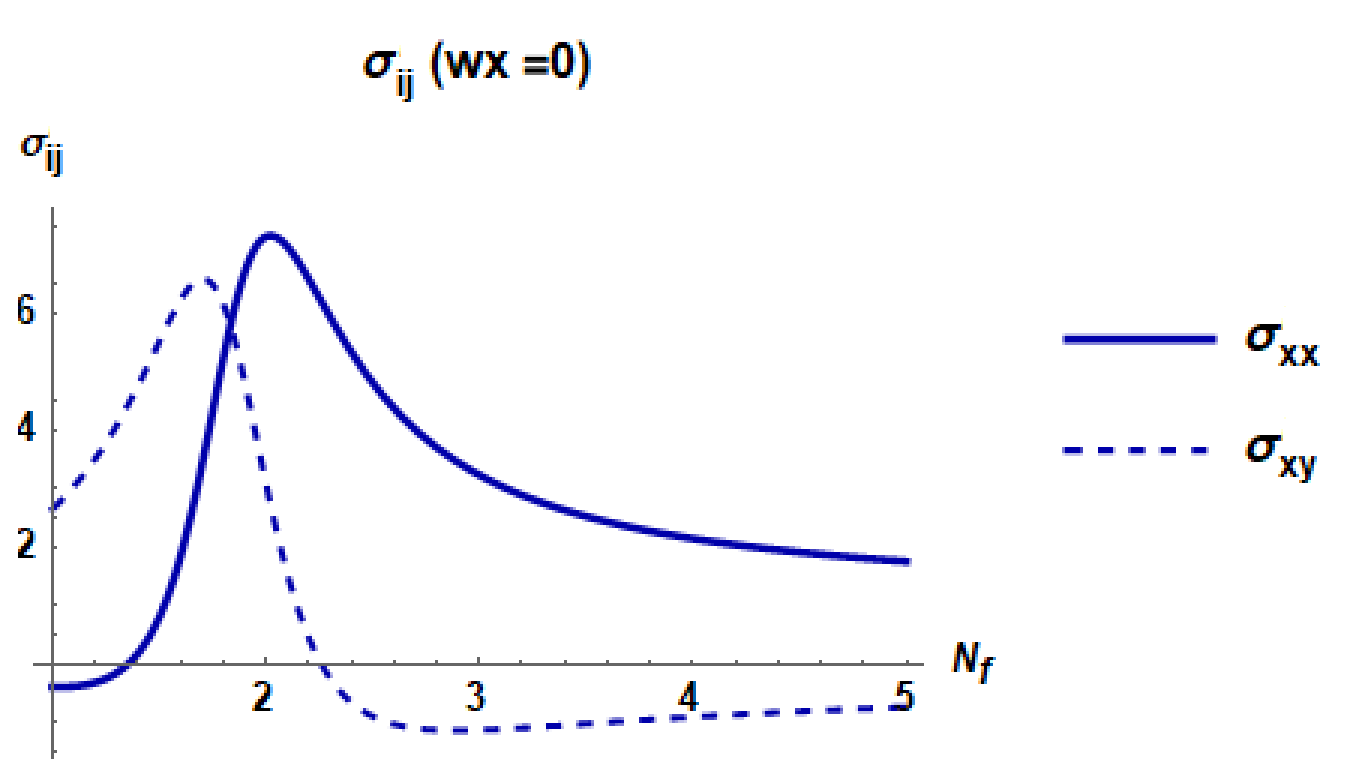}
  \caption{Zero Coulomb interaction $(\wx=0)$ and finite disorder}
  \label{fig:sub1}
\end{subfigure}%
\begin{subfigure}{.3\textwidth}
  \centering
  \includegraphics[scale=0.4]{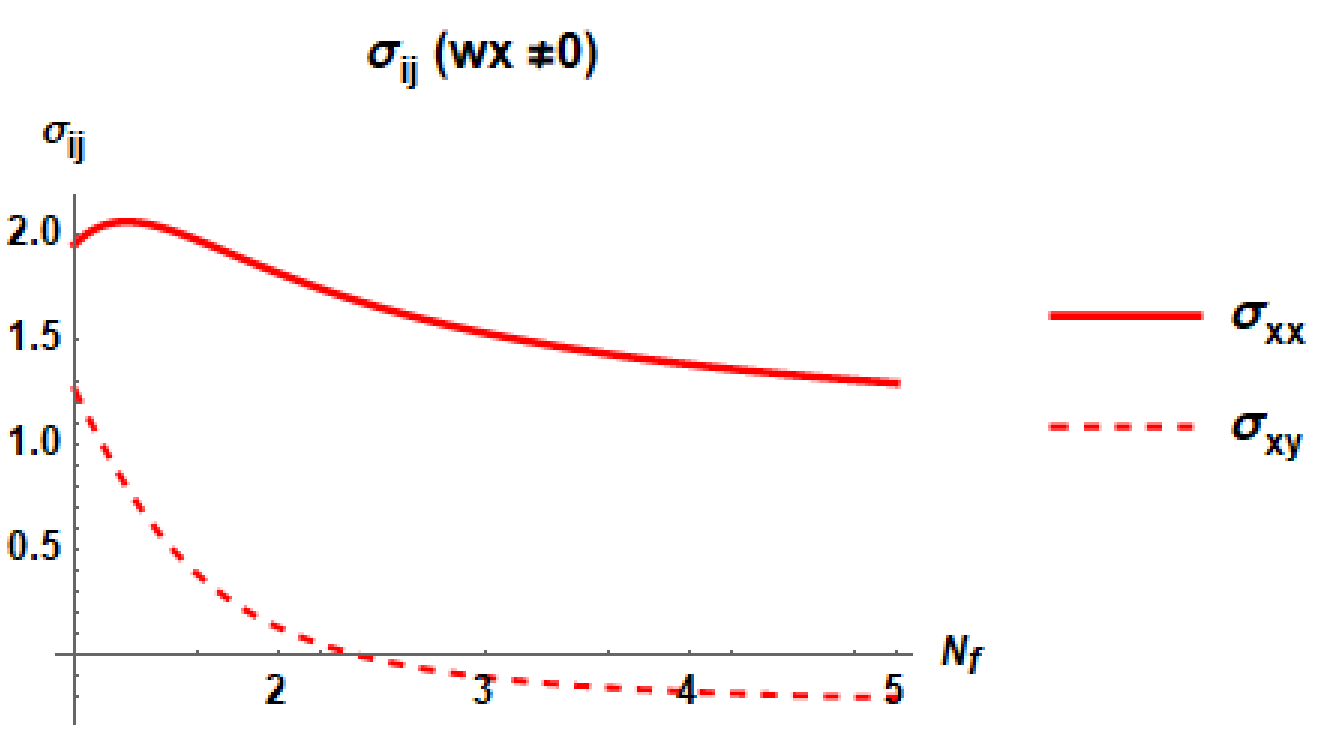}
  \caption{Finite Coulomb interaction $(\wx\neq 0)$ and finite disorder}
  \label{fig:sub2}
\end{subfigure}%
\begin{subfigure}{.3\textwidth}
  \centering
  \includegraphics[scale=0.4]{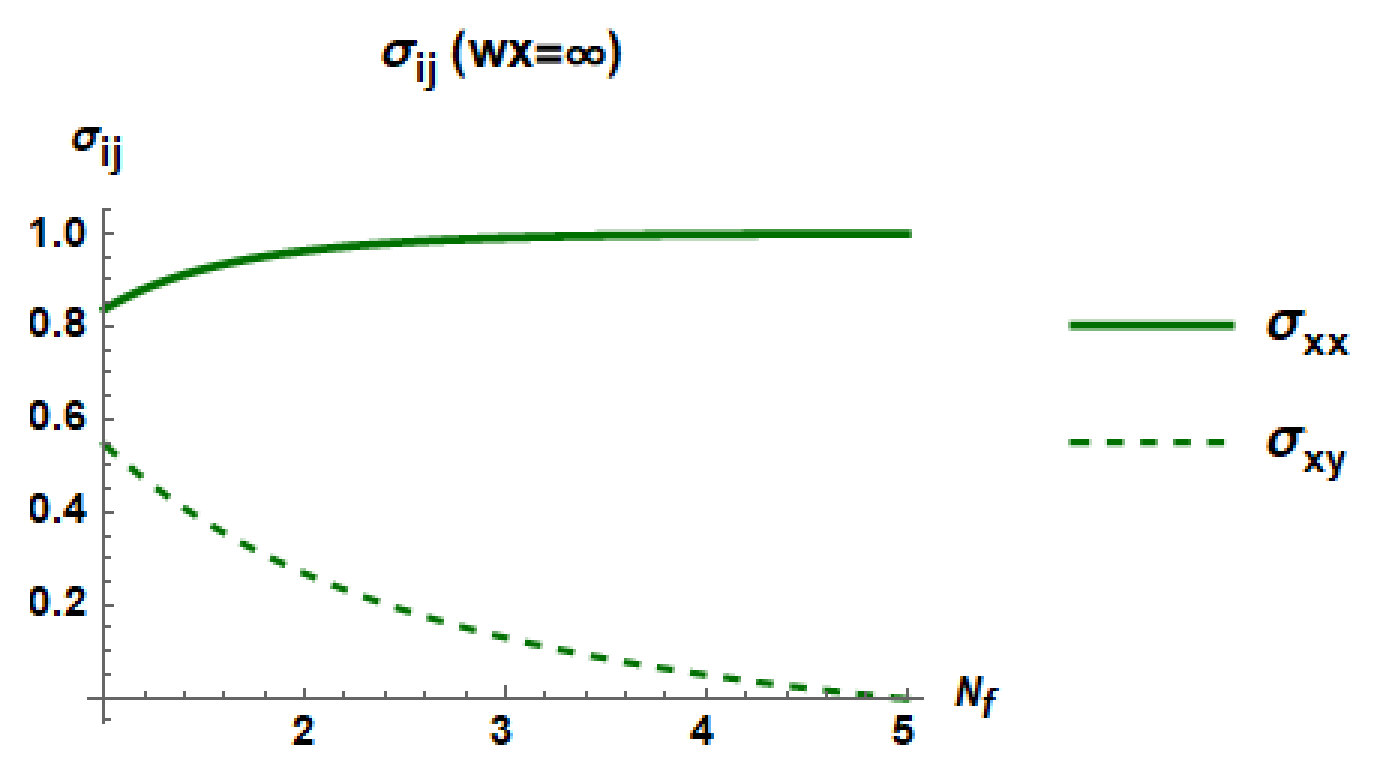}
  \caption{Infinite Coulomb interaction $(\wx=\infty)$ without disorder}
  \label{fig:sub3}
\end{subfigure}%
\caption{Electrical response components for two fixed points with varying number of fermion flavor $N_f$. The values in the figures are all in the unit of $\frac{1}{2\pi}$ }
\label{fig:Electric-Sigmaij-Nf}
\end{figure}

We now use \eqref{conduct-xx-relation} and \eqref{conduct-xy-relation} and the composite fermion conductivities computed in the previous section to obtain the electrical conductivities.
We begin with the fixed point without the Coulomb interaction.
Plugging in the two-loop composite fermion conductivities \eqref{sigma-xx-wx=0} and \eqref{sigma-xy-wx=0} into Eqs.~\eqref{conduct-xx-relation} and \eqref{conduct-xy-relation}, we find the dc electrical conductivities to ${\cal O}(1/N_f$):
\begin{align}
\label{nocoulombresult}
\big(\sigma_{xx}, \sigma_{xy} \big) = \Big(0.97 + {2.26 \over N_f}, - 0.24 - {1.91 \over N_f} \Big) \cdot {1 \over 2 \pi}, \quad \big(w_x = 0 \big).
\end{align}
Similarly, for the fixed point with Coulomb interaction, we use the two-loop composite fermion conductivities \eqref{sigma-xx-wx-Not=0} and \eqref{sigma-xy-wx-Not=0} to obtain the dc electrical conductivities:
\begin{align}
\label{withcoulombresult}
\big(\sigma_{xx}, \sigma_{xy} \big) = \Big(0.97 + {1.43 \over N_f}, - 0.24 - {0.06 \over N_f} \Big) \cdot  {1 \over 2 \pi} , \quad \big(w_x \neq 0 \big).
\end{align}
At ${\cal O}(1/N_f^0)$, we find that self-duality \eqref{pvsymmetryrelation} occurs with small, but nonzero $\sigma_{xy}$ and a longitudinal conductivity $\sigma_{xx} \lesssim R^{-1}_Q$.
Finite $1/N_f$ corrections, due to the fluctuating Chern-Simons gauge field and disorder, lead to violations of self-duality.
These violations are milder when the Coulomb interaction is included $(w_x \neq 0$) than when only short-ranged interactions are present ($w_x = 0$).
The expressions for the quantum critical electrical conductivities, \eqref{nocoulombresult} and \eqref{withcoulombresult}, are reliable at large $N_f$.
To interpolate these results to small, but finite $N_f$, we use the exact expressions that obtain from substituting the composite fermion conductivities, either with \eqref{sigma-xx-wx=0} and \eqref{sigma-xy-wx=0} or without \eqref{sigma-xx-wx-Not=0} and \eqref{sigma-xy-wx-Not=0} the Coulomb interaction, into Eqs.~\eqref{conduct-xx-relation} and \eqref{conduct-xy-relation}.
The results are plotted in Figs.~\ref{fig:sub1} and \ref{fig:sub2}.
As we saw before in \eqref{sigma-xx-wx=0}, Fig.~\ref{fig:sub1} indicates the extrapolation $N_f \rightarrow 1$ is dubious, since it results in a negative longitudinal conductivity.

We end this section with a speculative comment about another way to obtain self-duality.
Examining the composite fermion Hall conductivity \eqref{sigma-CF-xy}, we observe that $\sigma_{xy}^\psi \sim {1 \over \text{finite} + w_x}$, where ``finite" refers to an ${\cal O}(1)$ constant.
Taking $w_x \rightarrow \infty$, then results in $\sigma_{xy}^\psi = 0$.
We plot the associated electrical conductivities in Fig.~\ref{fig:sub3}.
We do not have an argument for why the Coulomb coupling would flow to strong coupling;
$w_x = \infty$ is not a fixed point of the beta functions \eqref{beta-wx} and \eqref{beta-gm}.
We note that in \cite{PhysRevResearch.2.023303} we studied, in addition, the effect of a dissipative Coulomb interaction, which allowed for fixed points with infinite strength Coulomb interaction.
The gauge field propagator at such fixed points is different from that considered in this paper and so a determination of the electrical conductivity at such fixed points is left for future work.

\section{Discussion}
\label{discussion}

In this paper, we studied the quantum critical electrical conductivity at the superconductor-insulator transition in two spatial dimensions.
We focused on transitions modeled by a theory of charge-2e bosons on a lattice at commensurate filling, with charge-conserving quenched disorder.
The critical properties are those of the dirty 3d XY universality class.
For our study, we used a composite fermion theory, consisting of a single Dirac fermion coupled to a Chern-Simons gauge field with quenched random mass disorder, that is dual to the dirty 3d XY model.
There are at least two reasons to prefer the composite fermion description.
The first is that it can exhibit a particle-vortex symmetry at the level of its Lagrangian.
A consequence of this symmetry is the semicircle law \eqref{pvsymmetryrelation}, which appears to be realized in some experiments \cite{Breznay2016}.
A second reason to prefer the composite fermion theory is that an unscreened Coulomb interaction is straightforwardly included in the model.
This allows for a comparison of the properties of the theory with or without the Coulomb interaction.

Our calculation of the critical conductivity was performed in an expansion in $1/N_f$, where $N_f$ is the number of fermion flavors.
Our results are summarized by Eqs.~\eqref{nocoulombresult} and \eqref{withcoulombresult} and in Figs.~\ref{fig:sub1} and \ref{fig:sub2}.
At order $1/N_f^0$, the theory is self-dual, realizing the semicircle law \eqref{pvsymmetryrelation} with $\sigma_{xx} \lesssim (2e)^2/h$ and small, finite $\sigma_{xy}$.
Fluctuation effects due to the gauge field and the random mass contribute corrections of order $1/N_f$ to the conductivity that violate self-duality.
We find these violations to be smaller when the Coulomb interaction is present.

In the previous section, based on the leading corrections to the composite fermion Hall conductivity, we speculated that infinite strength Coulomb interactions of the sort we studied in \cite{PhysRevResearch.2.023303} might produce self-dual transport.
We leave an explicit calculation of the critical conductivity at infinite strength Coulomb interactions for future work.
We note that such fixed points have dynamical critical exponent $z > 1$.

It would be interesting to carry out these electrical transport calculations at the dirty XY model fixed point found in \cite{PhysRevB.101.144506}, i.e., within the usual bosonic description, and to compare the results to those in this paper.
The fixed point in \cite{PhysRevB.101.144506} does not include the unscreened Coulomb interaction.
The bosonic description of the dirty XY model fixed point gives critical exponents that are in closer agreement with numerical experiment \cite{PhysRevB.94.134501} when $N_f \rightarrow 1$, than those of the composite fermion theory (at least to ${\cal O}(1/N_f)$).

There are two aspects as to why we focused on a theory corresponding to a model of lattice bosons with charge-conserving disorder. 
First, regarding the type of disorder within this lattice boson model, we found in \cite{PhysRevResearch.2.023303} that other types of disorder do not yield accessible disordered fixed points with or without a finite Coulomb interaction.
These calculations were performed in a $1/N_f$ expansion; it is possible that a different choice of artificial expansion parameter (perhaps one that preserves the same symmetries as the $N_f = 1$ theory) may find nontrivial fixed points.
Second, regarding the commensurability constraint, the dual composite fermions acquire a finite chemical potential if this is relaxed. 
It would be extremely interesting to find disordered fixed points in this more general situation, as we believe it more closely resembles the experimentally-realized superconductor-insulator transitions.
Recent work has shown \cite{10.21468/SciPostPhys.13.5.102, 10.21468/SciPostPhys.14.5.113} how the optical conductivity is constrained by the anomaly structure of emergent symmetries in a class of non-Fermi liquids (so-called ``Hertz-Millis" theories) consisting of a Fermi surface coupled to a gapless bosonic order parameter.
Such models are closely related to the theory we studied (see \cite{2022arXiv220904472M} for a related theory studied recently), when the composite fermion density is nonzero.
It would be worthwhile to understand the interplay of this emergent anomaly structure with particle-vortex symmetry.

\section*{Acknowledgements}

We thank Hart Goldman, Yen-Wen Lu, and Sri Raghu for useful discussions. 
This material is based upon work supported by the U.S. Department of Energy, Office of Science, Office of Basic Energy Sciences under Award No.~DE-SC0020007. 

\appendix
\section{Gauge boson propagator} \label{appendix-gauge-propagator} 

Start from Eq.~\eqref{Sgauge-propagator}, with 
$A \equiv \frac{ +\gX^2   \bm{k}^2  }{ k   } $,
$ B \equiv  i \kappa \, |\bm{k}|$, and
$C \equiv g_X^2  \, k  + |\bm{k}| \,\wx $.
This one-loop effective action includes the leading-order polarization:
\begin{eqnarray}
&&
\Pi_{00} = 
-\gX^2
\frac{    \bm{k}^2  }{  k   }  , \;\;\;       \\
&&
\Pi_{TT}
=
(\delta_{ij} - \frac{k_i k_j}{\bm{k}^2})\Pi_{ij}
=
-\gX^2  \, k.
\end{eqnarray}
We work in Coulomb gauge (vanishing longitudinal component of the gauge field).
Including the gauge-fixing term, $\int_x \frac{(\nabla_j a_j)^2 }{2\lambda}$, which preserves $SO(2)$ rotation invariance, the one-loop effective action is
\begin{eqnarray}
&&
S_{\rm gauge}^{(\text{gauge-fixed})} \equiv  \frac{1}{2} \int_k
\begin{pmatrix}
a_0 &  a_T
\end{pmatrix}_k
\begin{pmatrix}
A &   B  \\
B  &C \\
\end{pmatrix}
\begin{pmatrix}
a_0 \\  a_T
\end{pmatrix}_{-k}       \\
  &&
 =
 \lim_{\lambda \rightarrow 0}
  \frac{1}{2} \int_k
\begin{pmatrix}
a_0 & a_x  & a_y
\end{pmatrix}_{k}
(
\begin{pmatrix}
A & B  \frac{i k_y }{|\bm{k}|} & B  \frac{-i k_x }{|\bm{k}|}  \\
B  \frac{-i k_y }{|\bm{k}|} &  \frac{ k_y^2 }{|\bm{k}|^2}C & \frac{ - k_x k_y }{|\bm{k}|^2}C\\
B \frac{+ i k_x }{|\bm{k}|} & \frac{ - k_x k_y }{|\bm{k}|^2} C& \frac{ k_x^2 }{|\bm{k}|^2} C
\end{pmatrix} 
+
\frac{1}{\lambda}
\begin{pmatrix}
0 & 0 & 0 \\
0 & k_x k_x &  k_x k_y \\
0 & k_x k_y  & k_y k_y
\end{pmatrix} 
)
\begin{pmatrix}
a_0 \\
a_x \\
a_y
\end{pmatrix}_{-k},
\end{eqnarray}
where $\int_k \equiv \int \frac{d^3k}{(2\pi)^3}$.
We obtain the gauge field propagators \eqref{D00} - \eqref{Dij}, upon first inverting the $3 \times 3$ matrix kernel and then taking the limit $\lambda \rightarrow 0$.

\section{Random mass correction to the composite fermion conductivity} \label{appendix-random-mass-correction}

In this appendix, we detail the evaluation of Eqs.~\eqref{Pi-a-mass} and \eqref{Pi-bc-mass}, following \cite{PhysRevB.95.235146}.

\subsection{ $\PiM^{(b+c )}$  }

We first consider the subdiagram of Eq.~\ref{Pi-bc-mass}:
\begin{eqnarray}
&&
 \begin{tabular}{lll}
   &  \includegraphics[scale=0.5]{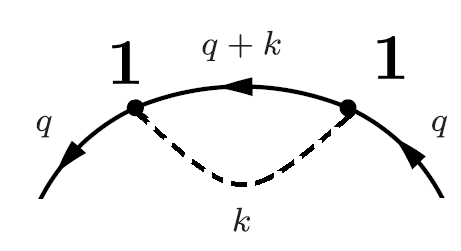}    &
\end{tabular}
=
\int  \frac{d^3 k}{(2\pi)^3}
\bm{1}   \gamma_\sigma  \, \bm{1}
(+i) 
\frac{ \, ( q+k)_\sigma  }{ (q_0 +k_0)^2 + \bm{(k+q)}^2} 
2\pi \delta(k_0)  \\
&&
=
\int  \frac{d^2 k}{(2\pi)^2}
  \gamma_\sigma  
(+i)
\frac{ \, (q_0   ) \delta_{\sigma 0} + (q_a+k_a) \delta_{\sigma a}   }
{ ( q_0 +0)^2 + (\bm{k+q})^2}    \\
&&
=
\int  \frac{d^2 \bm{\ell}}{(2\pi)^2}
  \gamma_\sigma  
(+i)
\frac{ \, (q_0   ) \delta_{\sigma 0} + (\ell_a) \delta_{\sigma a}   }
{ ( q_0 +0)^2 + (\bm{\ell})^2}  
\equiv 
i \, \gamma_0 \, I_X(q_0 )  \, \delta_{\sigma 0} +0 , 
\end{eqnarray}
Then, the full diagram of  Eq.~\ref{Pi-bc-mass} reads: 
\begin{eqnarray}
&&  
\Pi^{(b+c)}_{xx} (p_0,\bm{p}=0 )  \\
&&
=- 2 (\frac{ig}{\sqrt{N_f}} )^{2} \,N_f\, 
\int_q
\Tr[ \gamma_x G(q) 
\Big( \int \frac{d^3k}{(2\pi)^3} \bm{1}  G(k+q)  \bm{1} \times  2\pi \delta(k_0) g_m \; \Big)
G(q) \gamma_x  G(p+q)   ] \,    \qquad \\
&&
 =- 2 (\frac{ig}{\sqrt{N_f}} )^{2} \,N_f\, 
\int_q
\Tr[ \gamma_x G(q) 
 \Big( i \, I_X \gamma_0  + 0  \Big) 
G(q) \gamma_x  G(p+q)   ] 
\times  g_m\,  \\
&&
=
- 2 (\frac{ig}{\sqrt{N_f}} )^{2} \,N_f\, 
\int_q
\Tr[ \gamma_x   \frac{ (+i) q_\rho \gamma_\rho  }{ q^2_0 + \bm{q}^2  }
\Big( i \, I_X \gamma_0  + 0 \Big)
\frac{ (i) q_\tau \gamma_\tau  }{q^2_0 + \bm{q}^2  } 
\gamma_x 
(i) \frac{ (q  +p)_{\kappa} \gamma_\kappa }{ (q_0+ p_0)^2 + (\bm{q+0})^2}   ] 
\times    g_m. \, \quad \quad \,
\end{eqnarray}
Using
\begin{eqnarray}
\Tr[ \gamma_x  \gamma_\rho \gamma_0  	\gamma_\tau \gamma_x \gamma_\kappa] 
\, q_\rho \, q_\tau   \, (q+p)_\kappa  
= 2 (p_0+q_0) \, (\bm{q^2} -q_0^2) 
\end{eqnarray}
we obtain 
\begin{eqnarray}
&&
\Pi^{(b+c)}_{xx} (p_0,\bm{p}=0 )
=
 2 (-1) (\frac{+ig}{\sqrt{N_f}} )^{2} \,N_f\,  g_m 
\int \frac{d^2 \bm{q} dq_0}{(2\pi)^3}
\frac{1}{ [q_0^2+\bm{q}]^2} 
\frac{ 2 (p_0+q_0) \, (\bm{q^2} -q_0^2)   }{ (q_0 + p_0)^2+ \bm{q}^2 }
 I_X(q_0 )  \cr
 &&
 =
  2 (-1) (\frac{+ig}{\sqrt{N_f}} )^{2} \,N_f\,  g_m 
\int \frac{d^2 \bm{q} dq_0}{(2\pi)^3}
\frac{ 2 (p_0+q_0) \, (\bm{q^2} -q_0^2)     }{ [q_0^2+\bm{q}]^2 \;[(q_0 + p_0)^2+ \bm{q}^2]    } 
\int^{\Lambda}  \frac{d^2 \bm{\ell}}{(2\pi)^2}
\frac{ q_0 }
{ q_0^2 +  \bm{\ell} ^2}  
\label{Pibc-gm-simp}  \cr
&&
=
  2 (-1) (\frac{+ig}{\sqrt{N_f}} )^{2} \,N_f\,  g_m    \int \frac{ dq_0}{ 2\pi  }
\Big(
2 (p_0+q_0)q_0 \nn \\
&& \times
\frac{1}{4\pi  }
\frac{ 2 q_0^2 - 2 (p_0+q_0)^2  + [(p_0+q_0)^2 +q_0^2] \log[\frac{(q_0+p_0)^2}{q_0^2}]  }
{ [ (p_0+q_0)^2 -q_0^2 ]^2} 
\Big)
\frac{1}{4\pi}   \log[\frac{\Lambda^2}{q_0^2}].  \qquad \;
\label{Exist-Lambda-expression}
\end{eqnarray}
Here, we used 
\begin{eqnarray}
\int_0^{\Lambda } \frac{ d^2 \ell }{(2\pi)^2}  
\frac{1}{q_0^2 + \ell^2}
=
\frac{1}{4\pi} 
\log[ \frac{\Lambda^2 + q_0^2 }{q_0^2}]
\to 
\frac{1}{4\pi}   \log[\frac{\Lambda^2}{q_0^2}].
\end{eqnarray}
We simplify the integrand by defining 
\begin{eqnarray}
&& K( q_0) \cr
\, && \equiv
\Big(
2 (p_0+q_0)q_0
\times
\frac{1}{4\pi  }
\frac{ 2 q_0^2 - 2 (p_0+q_0)^2  + [(p_0+q_0)^2 +q_0^2] \log[\frac{(q_0+p_0)^2}{q_0^2}]  }
{ [ (p_0+q_0)^2 -q_0^2 ]^2} 
\Big) 
\frac{1}{4\pi}   ( -\log[q_0^2] )  \quad  \qquad \;  \cr
&&
=
( \frac{ - \log[q_0^2] }{4\pi}    ) 
\frac{  (p_0+q_0)q_0  }{  \pi  (p_0+2q_0) p_0 }
\Big(
-1
+
\frac{1}{2}
\frac{   [(p_0+q_0)^2 +q_0^2] \log[\frac{(q_0+p_0)^2}{q_0^2}]  }
{   (p_0+2q_0)^2 \, p_0^2   } 
\Big) 
\end{eqnarray}
A change of variables gives 
\begin{eqnarray}
&&
\int_{-\infty}^{\infty} \frac{dq_0}{2\pi }  K(q_0  = x p_0 ) \cr
&& =
|p_0|
  \int_{-\infty}^{\infty} \frac{dx}{2\pi} \, 
 \frac{-1    }{4\pi}  (\log[ x^2] + \log[p_0^2])
 \frac{ x (1 + x)}{ 1 (1 + 2 x)}
\,
\frac{-1}{ \pi}
\Big[
 1   +   \frac{ x^2 +( x+1)^2 }{2   (1+ 2 x) }
 \log [ \frac{x^2}{( x+1)^2}]    
\;   \Big] \quad   \cr
&&
=
|p_0|
  \int_{-\infty}^{\infty} \frac{dx}{2\pi} \, 
 \frac{-1    }{4\pi}  (\log[ x^2]  )
 \frac{ x (1 + x)}{ 1 (1 + 2 x)}
\,
\frac{-1}{ \pi}
\Big[
 1   +   \frac{ x^2 +( x+1)^2 }{2   (1+ 2 x) }
 \log [ \frac{x^2}{( x+1)^2}]    
\;   \Big]    \nn   \\
&& \quad
+ |p_0|
  \int_{-\infty}^{\infty} \frac{dx}{2\pi} \, 
 \frac{-1    }{4\pi}  ( \log[p_0^2])
 \frac{ x (1 + x)}{ 1 (1 + 2 x)}
\,
\frac{-1}{ \pi}
\Big[
 1   +   \frac{ x^2 +( x+1)^2 }{2   (1+ 2 x) }
 \log [ \frac{x^2}{( x+1)^2}]  .
 \label{Kq0-sep}
\end{eqnarray}
The principal value of the second term, $\int dx \log[p_0^2] \times [...]$ is zero.
If we make the plot, we find it is an odd function with respect to $x=\frac{1}{2}$, so after we shift $x\to x-\frac{1}{2}$, the integral vanishes.
For the same reason, the $\log \Lambda^2$ term in Eq.~\eqref{Exist-Lambda-expression} vanishes.

Hence, we symmetrize the first line of Eq.~\ref{Kq0-sep}, the $\log q_0^2$ term of Eq.~\eqref{Exist-Lambda-expression}:
\begin{eqnarray}
&&
\int_{-\infty}^{\infty} \frac{dq_0}{2\pi }  \frac{K(q_0   )+ K(-q_0   )}{2}
=
|p_0|
\int_{-\infty}^{\infty} \frac{dx}{2\pi } 
\frac{\log [x^2]}{4\pi^2}
\Big[
 \frac{ x (1 + x)}{ 1 (1 + 2 x)}
(  1   +   \frac{ x^2 +( x+1)^2 }{2   (1+ 2 x) }    \log [ \frac{x^2}{( x+1)^2}]   \nn \\
&& \qquad  \qquad \qquad \qquad \qquad \qquad
+
 \frac{ -x (1 - x)}{ 1 (1 - 2 x)}
(  1   +   \frac{ x^2 +( -x+1)^2 }{2   (1- 2 x) }    \log [ \frac{x^2}{( -x+1)^2}]    
\;  )
\Big]  \\
&&
=
\frac{1}{2\pi} \times \frac{ |p_0|  }{96}.
\end{eqnarray}
Thus, we conclude that
\begin{eqnarray}
&&
\Pi^{(b+c)}_{xx} (p_0,\bm{p}=0 )
=
 2 (-1) (\frac{+ig}{\sqrt{N_f}} )^{2} \,N_f\,  g_m 
 \times
 \frac{1}{2\pi} \times \frac{ |p_0|  }{96} .
\end{eqnarray}

\subsection{$\PiM^{(a )}$ }

Next we turn to Eq.~\eqref{Pi-a-mass}.
Again, consider the subdiagram,
\begin{eqnarray}
&&
 \begin{tabular}{lll}
  &  \includegraphics[scale=0.5]{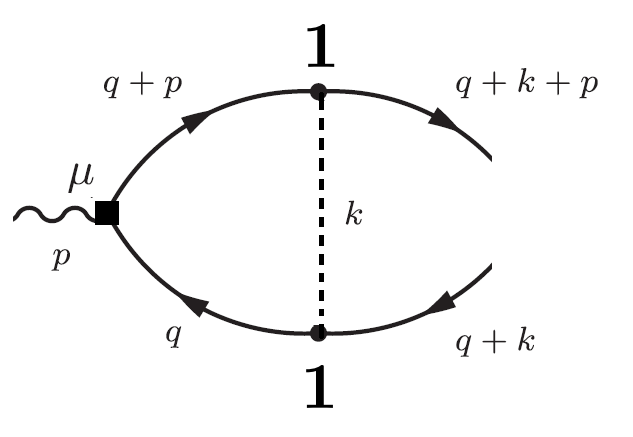}    &
\end{tabular}
=
\int \frac{d^3k}{(2\pi)^3}
i\frac{ (q_0+k_0+p_0  )\gamma_0+ \bm{(q+k+p) } \cdot  \vec{\gamma} }
{ (q_0+k_0+p_0  )^2 + (\bm{q+k+p})^2  }
\bm{1} \,
i\frac{ (q_0+ p_0  )\gamma_0+ \bm{(q+ p)} \cdot  \vec{\gamma} }
{ (q_0+ p_0  )^2 + (\bm{q+p})^2  }  \nn \\
&&
\qquad \qquad  \qquad \qquad  
\times  
\gamma_\mu \,
i\frac{ (q_0    )\gamma_0+ \bm{(q ) } \cdot \vec{\gamma} }
{ (q_0   )^2 + (\bm{q })^2  } 
\bm{1} \,
i\frac{ (q_0+k_0  )\gamma_0+ \bm{ (q+k )} \cdot  \vec{\gamma} }
{ (q_0+k_0 )^2 + (\bm{q+k })^2  }
\times 
2\pi \delta(k_0) \,g_m  \\
&&
= g_m 
\int \frac{d^2\bm{k}}{(2\pi)^2}
\gamma_\tau  \bm{1} \gamma_\kappa \gamma_\mu \gamma_\rho \bm{1}  \gamma_\sigma
\frac{ (q+k+p)_\tau  }{(q+k+p)^2}
\frac{ (q+ p)_\kappa }{(q+ p)^2}
\frac{ q_\rho }{  q^2 }
\frac{ (q+k)_\sigma }{ (q+k)^2}   \Big{|}_{k_0 =0} .
\end{eqnarray}
Consider the terms involving $k$.
Setting $\bm{p}=0$, we find:
\begin{eqnarray}
&&
I_{\tau \sigma} (q_0 ,p_0)
\equiv 
\int \frac{d^2\bm{k}}{(2\pi)^2}
\frac{ (q+k+p)_\tau  }{(q+k+p)^2}
\frac{ (q+k)_\sigma }{ (q+k)^2}   \Big{|}_{k_0 =0, \bm{p} =0 }    \\
&&
=
\int \frac{d^2\bm{\ell}}{(2\pi)^2}
\frac{ (q_0+k_0+p_0) \delta_{\tau 0} + (\bm{\ell+p})_\tau \delta_{\tau a}   }
{ (q_0+k_0+p_0)^2 + \bm{\ell}^2   }
\; 
\frac{ (q_0+k_0 ) \delta_{\sigma 0} + (\bm{\ell })_\sigma \delta_{\sigma b} }
{ (q_0+k_0 )^2 + \bm{\ell}^2 }
 \Big{|}_{k_0 =0, \bm{p} =0 }   \\
 &&
= 
\int \frac{d^2\bm{\ell}}{(2\pi)^2}
\frac{ (q_0 +p_0) \delta_{\tau 0} + (\bm{\ell })_\tau \delta_{\tau a}   }
{ (q_0+ p_0)^2 + \bm{\ell}^2   }
\; 
\frac{ (q_0+0) \delta_{\sigma 0} + (\bm{\ell })_\sigma \delta_{\sigma b} }
{ (q_0+ 0 )^2 + \bm{\ell}^2 }   \\
&&
=
\int \frac{d^2\bm{\ell}}{(2\pi)^2}
\frac{ (q_0+p_0)q_0 \,  \delta_{\tau 0} \delta_{\sigma 0} }
{ [(q_0+ p_0)^2 + \bm{\ell}^2 ] \, [q_0^2+ \bm{\ell}^2   ]   }
+
\int \frac{d^2\bm{\ell}}{(2\pi)^2}
\frac{ \bm{\ell}^2  \delta_{ab} \delta_{\tau a} \delta_{\sigma b}  }{d}
\frac{1}{ [(q_0+ p_0)^2 + \bm{\ell}^2 ] \, [q_0^2+ \bm{\ell}^2   ]  } \quad \quad
\end{eqnarray}

Now we set $\mu =\nu = x$ .
First consider the temporal component of the gamma index trace in Eq.~\eqref{Pi-a-mass}:
\begin{eqnarray}
\Tr[ \gamma_\tau    \gamma_\kappa \gamma_{\mu=x} \gamma_\rho   
\gamma_\sigma \gamma_{\nu=x}] \,  \delta_{\tau 0} \delta_{\sigma 0}
=
\Tr[ \gamma_0    \gamma_\kappa \gamma_{x} \gamma_\rho   
\gamma_0 \gamma_{x}]
=
(-1) \Tr[    \gamma_\kappa \gamma_{x} \gamma_\rho     \gamma_{x}  ].
\end{eqnarray}
Contracting this with the momentum gives
\begin{eqnarray}
(-1) \Tr[    \gamma_\kappa \gamma_{x} \gamma_\rho     \gamma_{x}  ]
\, (q+ p)_\kappa   q_\rho 
=
2 q_0 (p_0+q_0 ).
\end{eqnarray}

Next consider the spatial components of the gamma index trace.
We perform the trace in $2-\epsilon$ spatial dimensions, rather than $2$ spatial dimensions:
\begin{eqnarray}
&&
\Tr[ \gamma_\tau    \gamma_\kappa \gamma_{\mu=x} \gamma_\rho   
\gamma_\sigma \gamma_{\nu=x}] \,  
                 \delta_{ab} \delta_{\tau a} \delta_{\sigma b}    \cr
&&                              
=   
\sum_{a = 2-\epsilon \text{~number of indices}  }
\Tr[ \gamma_a    \gamma_\kappa \gamma_{x} \gamma_\rho   
\gamma_a \gamma_{x}]  
=
\Tr[   \gamma_\kappa \gamma_{x} \gamma_\rho   
\sum_{a } (\gamma_a \gamma_{x}  \gamma_a) ]   \\
&&
=
  \Tr[   \gamma_\kappa \gamma_{x} \gamma_\rho   
 (2\gamma_x -(2-\epsilon)\gamma_x  ) ]    
= \epsilon  \;    \Tr[   \gamma_\kappa \gamma_{x} \gamma_\rho  \gamma_x ].
\end{eqnarray}
Contracting this with the momentum yields
\begin{eqnarray}
\epsilon  \;    \Tr[   \gamma_\kappa \gamma_{x} \gamma_\rho  \gamma_x ]
\, (q+ p)_\kappa   q_\rho 
=
 \epsilon \times (- 2) q_0 (p_0+q_0 ).
\end{eqnarray}

Putting this all together, we find:
\begin{eqnarray}
&&
\Pi^{(a)}_{\mu \nu} (p_0, \bm{p}=0)
= (-1) (\frac{+ig}{\sqrt{N_f}} )^{2}\,N_f\, g_m
\int \frac{d^3q}{(2\pi)^3}
\frac{ 2 q_0 (p_0+q_0 ) }
{[(q_0+ p_0)^2 + (\bm{q+ 0} )^2  ]\; [q_0^2+ \bm{q}^2]}   \nn \\
&&  \times
\Big(
\int \frac{d^2\bm{\ell}}{(2\pi)^2}
\frac{ (q_0+p_0)q_0 \,    }
{ [(q_0+ p_0)^2 + \bm{\ell}^2 ] \, [q_0^2+ \bm{\ell}^2   ]   } 
- \epsilon
\int \frac{d^{2-\epsilon}\bm{\ell}}{(2\pi)^2}
\frac{ \bm{\ell}^2   }{2}
\frac{1}{ [(q_0+ p_0)^2 + \bm{\ell}^2 ] \, [q_0^2+ \bm{\ell}^2   ]  }
\Big).
\label{Pia-gm-simp}
\end{eqnarray}
We wish to extract finite parts of this expression.
The finite part of the first $d^2 \bm{\ell}$ integral is
\begin{eqnarray}
\int \frac{d^2\bm{\ell}}{(2\pi)^2}
\frac{ (q_0+p_0)q_0 \,    }
{ [(q_0+ p_0)^2 + \bm{\ell}^2 ] \, [q_0^2+ \bm{\ell}^2   ]   } 
=
 \frac{1}{4\pi}
\frac{ (q_0+p_0)q_0  }{  p_0 (p_0 + 2 q_0)  }
\log[\frac{(q_0+ p_0)^2}{q_0^2 }].
\end{eqnarray}
For the second integral, we need to extract the $\frac{1}{\epsilon}$ part. 
Using the usual Feynman parametrization,
\begin{eqnarray}
&&
\int \frac{d^{2-\epsilon}\bm{\ell}}{(2\pi)^2}
\frac{ \bm{\ell}^2   }{2}
\frac{1}{ [(q_0+ p_0)^2 + \bm{\ell}^2 ] \, [q_0^2+ \bm{\ell}^2   ]  }
=
\int \frac{d^{2-\epsilon}\bm{\ell}}{(2\pi)^2}
\int_0^1 dx 
\frac{ 1   }{2}
\frac{ \bm{\ell}^2 }{  [ \bm{\ell}^2 + \Delta(x,q_0,p_0)]^2}  \\
&&
=
\int_0^1 dx  
\frac{1}{2}  
\frac{1}{2\pi \epsilon} + \mathcal{O}(\epsilon^0)
=
\frac{1}{4 \pi \epsilon} + \mathcal{O}(\epsilon^0).
\end{eqnarray}
It remains to perform the $d^2 \bm{q}$ integral, which has the same structure as the one above:
\begin{eqnarray}
\int \frac{d^3q}{(2\pi)^3}
\frac{ 2 q_0 (p_0+q_0 ) }
{[(q_0+ p_0)^2 +  \bm{q } ^2  ]\; [q_0^2+ \bm{q}^2]} 
=
\int \frac{dq_0}{2\pi}
 \frac{1}{2\pi}
\frac{ (q_0+p_0)q_0  }{  p_0 (p_0 + 2 q_0)  }
\log[\frac{(q_0+ p_0)^2}{q_0^2 }].
\end{eqnarray}
Putting this together, we have
\begin{eqnarray}
&&
\Pi^{(a)}_{\mu \nu} (p_0, \bm{p}=0)
= (-1) (\frac{+ig}{\sqrt{N_f}} )^{2}\,N_f\, g_m
\int_{-\infty}^{\infty} \frac{dq_0}{2\pi}
 \frac{1}{2\pi}
\frac{ (q_0+p_0)q_0  }{  p_0 (p_0 + 2 q_0)  }
\log[\frac{(q_0+ p_0)^2}{q_0^2 }]   \nn \\
&&
\qquad \qquad\qquad\qquad\qquad\qquad\qquad\qquad 
\times 
\Big(
 \frac{1}{4\pi}
\frac{ (q_0+p_0)q_0  }{  p_0 (p_0 + 2 q_0)  }
\log[\frac{(q_0+ p_0)^2}{q_0^2 }]
-
\frac{1}{4\pi }
\Big)  \\
&&
=
(-1) (\frac{+ig}{\sqrt{N_f}} )^{2}\,N_f\, g_m
\times \frac{1}{2\pi}
\times  \frac{|p_0|}{96}.
\end{eqnarray}
 
Summarizing the results of this section, we find the random mass diagrams equal
\begin{eqnarray}
\Pi^{(a+b+c)}_{\mu \nu} (p_0, \bm{p}=0)
=
3 \times 
(-1) (\frac{+ig}{\sqrt{N_f}} )^{2}\,N_f\, g_m
\times \frac{1}{2\pi}
\times  \frac{|p_0|}{96}.
\end{eqnarray}
Notice there is no off-diagonal component.
By Eq.~\eqref{sigma-ij-conductivity}, this self-energy contributes the following to the composite fermion conductivity:
\begin{eqnarray}
\sigma^{g_m}_{xx}  = \lim_{p_0 \to 0}  \frac{ \Pi^{(\text{S})} (\bm{p} = 0)  }{p_0} 
=
\frac{i}{\omega} \Pi^{xx} (\bm{p} = 0, |p_0| \to i |\omega|) 
=
  \frac{ - 3 g_m}{96 \times 2\pi},
\end{eqnarray} 
where we set the $g^2 =1 $ and $\omega >0 $.

\section{Gauge field correction to the composite fermion conductivity}  \label{appendix-gauge-correction}

In this appendix, we detail the evaluation of Eqs.~\eqref{Pi-a-gauge} and \eqref{Pi-bc-gauge}.
 
\subsection{Useful integrals} \label{all-fomula-needed}

To evaluate gauge field corrections to the composite fermion conductivity, we need the formulas given in the next two subsections.
In the expressions below, $\Lambda$ is a momentum integral UV cutoff.

\subsubsection{Basic integral building blocks}  \label{loop-integral-formula}

The simplest integrals we will need are the following:
\begin{eqnarray}
&& \int \frac{ d^3 k}{(2\pi)^3} \frac{1}{k^2} 
= \int_0^\Lambda  dk \frac{4\pi k^2 dk}{(2\pi)^3 \, k^2}
= \frac{\Lambda}{2 \pi^2} ,   \\
&& \int \frac{ d^3 k}{(2\pi)^3}  \frac{1}{k^2  (k+p)^2}
= \frac{1}{8 |p|}  ,    \\
 J_2 (p,q) \equiv
 &&  \int \frac{ d^3 k}{(2\pi)^3}  \frac{1}{(k+p)^2  (k+q)^2}
=\int \frac{ d^3 K}{(2\pi)^3}  \frac{1}{(K)^2  (K-p+q)^2}
= \frac{1}{8|p-q|} , \\
 J_3 (p,q) \equiv
 &&
\int \frac{ d^3 k}{(2\pi)^3}  \frac{1}{k^2 (k+p)^2  (k+q)^2}
= \frac{1}{ 8 |p|\, |q| \, |p-q| }   ,    \\
&&
\int \frac{ d^3 k}{(2\pi)^3}  \frac{k^2}{(k+p)^2  (k+q)^2}
= \frac{\Lambda }{2 \pi^2} + \frac{ p \cdot q }{8 |p-q|}.
\end{eqnarray}

Next we consider the integral, 
\begin{eqnarray}
I(p,q) \, && \equiv \int \frac{ d^3 k}{(2\pi)^3} \frac{k^2}{  (k+p)^2 (k+q)^2 (k+p+q)^2} \cr
&& =
 \int \frac{ d^3 k}{(2\pi)^3} 
 \frac{ (k+q)^2 +(k+p)^2-(k+p+q)^2 + 2 p \cdot q  }
 {  (k+p)^2 (k+q)^2 (k+p+q)^2}   \nn \\
&&
= \frac{1}{8 |q|}  + \frac{1}{ 8 |p|} 
- \frac{1}{8 |p-q|} 
+ \frac{2 p \cdot q}{8 |p||q| |p-q| }.
\end{eqnarray}
It is symmetric:  $I(p,q) = I(q,p)$
 
Next we consider the more complicated integral:
\begin{eqnarray}
  {\color{black}J_4(p,q)} \, && \equiv   \int \frac{d^3 k}{(2 \pi)^3}
\frac{1}{(k)^2 (k+p)^2 (k+q)^2 (k+p+q)^2} \cr
&& =
\frac{1}{8 |p| |q| (p \cdot q) }
\bigg( \frac{1}{|p-q|} -\frac{1}{|p+q|}  \bigg) .
\end{eqnarray} 
To prove this result, we re-express the integrand using partial fractions:
\begin{eqnarray}
&&  
 \int \frac{d^3 k}{(2 \pi)^3}
\frac{1}{(k)^2 (k+p)^2 (k+q)^2 (k+p+q)^2}   \\
&&
=
 \int \frac{d^3 k}{(2 \pi)^3}
\Big(
\frac{1}{(k)^2 (k+p)^2 (k+q)^2 } 
- \frac{1}{(k)^2 (k+p)^2  (k+p+q)^2}    \nn \\
&& \qquad  \qquad \qquad \qquad    
+
\frac{1}{  (k+q)^2   (k+p+q)^2 (k+p)^2 }
-
\frac{1}{  (k+q)^2   (k+p+q)^2 \, k^2 }
\Big) 
\frac{1}{2 p \cdot q}   \label{partial-integral-trick}    \\
&&
=
\frac{1}{8( p \cdot q ) \;|p|\, |q|  }
(\frac{1}{|p-q|} - \frac{1}{|p+q|}   )  . \qquad
\end{eqnarray}

Now consider the three-propagator integral
\begin{eqnarray}
L(p_5,p,k)
\, && \equiv 
 \int \frac{d^3 q}{(2 \pi)^3}
\frac{  q \cdot p_5 }{  (q)^2 (q+p)^2 (q+ k)^2 }  \cr
&& =
(-1)\frac{ (|k|+|p| - |k-p|  ) \; \Big(  |p| \,(p_5 \cdot k) + |k| \,(p_5 \cdot p)   \Big)  }
{16 \,|k|\, |p| \, (k\cdot p+ |k|\, |p|) \,|k-p|  }.
\label{3-propagator-linear-q}
\end{eqnarray}
 
Consider the four-propagator integral:
\begin{eqnarray}
 M(k ; p)
\equiv  \int \frac{d^3q}{(2\pi)^3}  
\frac{   \bm{ k \cdot q }   }{ q^2 (q+k)^2  (q+k+p)^2 (q+p)^2} .
\end{eqnarray}
$M(k ; p)$ is not symmetric with respect to $k,p$.
It is difficult to evaluate this four-propagator integral using the usual Feynman parameterization. 
(The three-propagator integral \eqref{3-propagator-linear-q} can be evaluated in this way.)
Instead, we use the partial fraction trick in Eq.~\eqref{partial-integral-trick} to rewrite it as multiple three-propagator integrals, and then use the 3-propagator result in Eq.~\eqref{3-propagator-linear-q}.
We find:
\begin{eqnarray}
&&
 M(k ; p)= 
 \int \frac{d^3q}{(2\pi)^3}  
\frac{   \bm{ k \cdot q }   }{ q^2 (q+k)^2  (q+k+p)^2 (q+p)^2}    \cr
&&
=
\int \frac{d^3q}{(2\pi)^3}
( \bm{ k \cdot q } )  
\Big(
\frac{  1  }{ q^2 (q+k)^2   (q+p)^2} 
-  \frac{1}{ q^2 (q+k)^2  (q+k+p)^2   }   \nn \cr
&&  \quad   
+ \frac{  1  }{  (q+k)^2  (q+k+p)^2 (q+p)^2} 
- \frac{1}{ q^2   (q+k+p)^2 (q+p)^2   }
\Big)
\frac{1}{2 (k \cdot p)}   \cr
&&
= \frac{1}{2 (k \cdot p)}
\Big(
L(\bm{k},k,p)  - L(\bm{k},k,k+p)
+ 0
- L(\bm{k},k+p,p)
\Big) \cr
&&
+  \frac{1}{2 (k \cdot p)}
\int \frac{d^3\ell}{(2\pi)^3}
\frac{  \bm{ (\ell- k) \cdot k } }{  \ell^2 (\ell+p)^2 (\ell-k +p)^2}   \cr
&&
=
\frac{1}{2 (k \cdot p)}
\Big(
L(\bm{k},k,p)  - L(\bm{k},k,k+p)
+   L(\bm{k},p,p-k)
-  \frac{ \bm{k}^2 }{ 8 |p|\, |p-k| \, |k| } \\
&&
- L(\bm{k},k+p,p)
\Big). \quad\quad \quad
\end{eqnarray}
 
A ``quadratic in $q$ integral" can be obtained from the above building blocks:
\begin{eqnarray}
&& \int \frac{d^3q}{(2\pi)^3}
\frac{  (k\cdot q) \,( \bm{ k \cdot q })  }{ q^2 (q+k)^2  (q+k+p)^2 (q+p)^2} \cr
&& = 
\int \frac{d^3q}{(2\pi)^3}
\frac{ \frac{1}{2}[ (q+k)^2 -k^2 -q^2 ]   }{ q^2 (q+k)^2  (q+k+p)^2 (q+p)^2} 
\,( \bm{ k \cdot q })    \qquad  \qquad  \\
&&
= \frac{1}{2}
\Big(
L(\bm{k} ,p,k+p)
- k^2 \,  M(k ; p)
-
L(\bm{k} , p, p-k)
+ \bm{k}^2  \frac{1}{ 8 |p| \, |p-k| \, |  k|   }
\Big).
\label{kq2D-kq3D}
\end{eqnarray}

\subsubsection{More complicated integral formulas}  
\label{loop-integral-formula-part2}

The momenta $p_5 = (p^0_5,p^x_5,p^y_5)$ and $p_5 = (q^0_6,q^x_6,q^y_6)$ appearing below are arbitrary external three-momenta that are independent of any integral momentum variables.
We list the following integrals:
\begin{eqnarray}
N(p_5; p,k ) \big{|}_{ p \neq k}
\, && \equiv
 \int \frac{d^3 q}{(2 \pi)^3}
\frac{  q \cdot p_5 }{  (q^2 ) \; (q^2) \; (q+p)^2 (q+ k)^2 } \cr
&& =
(-1) \frac{ |p|^2 (p_5 \cdot k) + |k|^2 ( p_5 \cdot p)  }{16 \, |k|^3 \,|p|^3  \,|k-p| } ,
\end{eqnarray}
\begin{eqnarray}
R_1 (p_5 ; k)
\equiv 
\int \frac{d^3q}{(2\pi)^3}
 \frac{ q \cdot p_5  }{q^2 (q+k)^2 q^2 }
 =
  (-1)  \frac{  p_5 \cdot k  \; }{16 \, |k|^3} ,
\end{eqnarray}
\begin{eqnarray}
R_2 (p_5, p_6  ; k)
\equiv 
\int \frac{d^3q}{(2\pi)^3}
 \frac{  (q \cdot p_5)  \;(q \cdot p_6)   }{q^2 (q+k)^2 q^2  } 
 = 
 \frac{ k^2 (p_5 \cdot p_6)  + (p_5 \cdot k) (p_6 \cdot k) }{32 \, |k|^3 } ,
\end{eqnarray}
 \begin{eqnarray}
S (p_5 ;  k,p)
\, && \equiv 
\int \frac{d^3q}{(2\pi)^3}
 \frac{ (q \cdot p_5)^2  }{q^2 (q+k)^2 \,  (q+p)^2} \cr
&& 
 =
 \frac{ 
   |p| \, (\, |k|+  2 |p| + |k-p| \,)  \,(p_5 \cdot k)^2
+   |k| \, (\, |p|+  2 |k| + |k-p| \,)  \,(p_5 \cdot p)^2
  }
 {16 \, |k-p| \; |k|  \;   |p|  \; \big(  \, |k|+|p| + |k-p| \,\big)^2 }  \nn \\
 && 
+
 \frac{ |k-p| \;|k| \; |p|  (\, |k|+|p| + |k-p| \,) \, p_5^2
+ 2 |k| \, |p| \,  (p_5 \cdot k) \, (p_5 \cdot p)
  }
 {16 \, |k-p| \; |k|  \;   |p|  \; \big(  \, |k|+|p| + |k-p| \,\big)^2 } ,
\end{eqnarray}
\begin{eqnarray}
R_3 (p_5 ; k,p)
\, && \equiv
\int \frac{d^3q}{(2\pi)^3}
 \frac{ (q \cdot p_5)^2  }{q^2 (q+k)^2 q^2 (q+p)^2}     \cr
&& 
=
\frac{ p^2  \,|k-p| \,k^2 \,p_5^2   \,(|k-p |  + |k| + |p|   )  
 +   2 k^2 \, p^2 \, (p_5 \cdot k) \, (p_5 \cdot p)   }
{16 \, |k-p|  \,|k|^3 \,  |p|^3 \; \big(  \, |k|+|p| + |k-p| \,\big)^2 }    \nn \\
&&
+
\frac{ k^3 (|k-p|+ |k| + 2|p|  ) \,(p_5 \cdot p)^2 
+     p^3 (|k-p|+ 2 |k| +  |p|  ) \,(p_5 \cdot k)^2 
}
{16 \, |k-p|  \,|k|^3 \,  |p|^3 \; \big(  \, |k|+|p| + |k-p| \,\big)^2 } .
\end{eqnarray}
Next we consider
\begin{eqnarray}
M_2 (p_5 ; k,p)
\, && \equiv
\int \frac{d^3q}{(2\pi)^3}
\frac{ (q \cdot p_5)^2   }{ q^2 (q+k)^2 (q+k+p)^2 (q+p)^2  }   \cr
&&
=
\int \frac{d^3q}{(2\pi)^3}
  (q \cdot p_5)^2   
\Big(
\frac{  1  }{ q^2 (q+k)^2   (q+p)^2} 
-  \frac{1}{ q^2 (q+k)^2  (q+k+p)^2   }   \nn \\
&& 
+ \frac{  1  }{  (q+k)^2  (q+k+p)^2 (q+p)^2} 
- \frac{1}{ q^2   (q+k+p)^2 (q+p)^2   }
\Big)
\frac{1}{2 (k \cdot p)}    \cr
&&
=
\frac{1}{2 (k \cdot p)} 
\Big[
 S (p_5 ;  k,p) 
-   S (p_5 ;  k, k+p)  \cr
&&  
+\int \frac{d^3q}{(2\pi)^3}  \frac{  (q \cdot p_5)^2    }{  (q+k)^2  (q+k+p)^2 (q+p)^2} 
- S (p_5 ;  k+p, p)  
\Big]  .
\end{eqnarray}
To further evaluate, we can shift the momentum:
\begin{eqnarray}
&& \int \frac{d^3q}{(2\pi)^3}
 \frac{  (q \cdot p_5)^2    }{  (q+k)^2  (q+k+p)^2 (q+p)^2} 
 =
 \int \frac{d^3\ell}{(2\pi)^3}
 \frac{  [ (\ell-k) \cdot p_5]^2    }{ \ell^2  (\ell+p)^2 (\ell-k + p)^2}     \nn \\
&&   
 =
  \int \frac{d^3\ell}{(2\pi)^3}
 \frac{  (\ell \cdot p_5 )^2 - 2 (\ell \cdot p_5) (k \cdot p_5) +   (k \cdot p_5)^2
     }{ \ell^2  (\ell+p)^2 (\ell+ p -k  )^2}     \nn \\
&&     
=
S(p_5; p,p-k)  - 2    (k \cdot p_5) \,L(p_5; p,p-k)
  +      (k \cdot p_5)^2            \frac{1}{8 |p|\, |p-k| \, |k|}.
\end{eqnarray}
Thus, we have reduced $M_2$ to a linear combination of integrals we have already computed:
\begin{eqnarray}
M_2 (p_5 ; k,p)
\, && =
\frac{1}{2 (k \cdot p)} 
\Big[
 S (p_5 ;  k,p) 
-   S (p_5 ;  k, k+p)  - S (p_5 ;  k+p, p)      \nn \\
&& 
+
S(p_5; p,p-k)  - 2    (k \cdot p_5) \,L(p_5; p,p-k)
  +      (k \cdot p_5)^2            \frac{1}{8 |p|\, |p-k| \, |k|}
\Big] .    \qquad  
\label{M2-Fun}
\end{eqnarray}

\subsection{ Anti-symmetric component of $ \Pi^{(b+c)}  $ }

The anti-symmetric part of the gauge field propagator is 
\begin{eqnarray}
D^{\text{Anti}}_{\alpha \beta }(k)
=
\frac{ \kappa   }{\kappa^2+ \gX^4    + \frac{ \wx  \gX^2 \,|\bm{k}|}{ |k|  }   }
(  \frac{  \epsilon_{ \alpha \beta \lambda} k_{\lambda}  \delta_{\lambda j}   }{ \bm{k}^2  } ) 
\equiv
\CAtd(k) \, 
 \frac{  \epsilon_{ \alpha \beta \lambda} k_{\lambda}  \delta_{\lambda j}   }{ \bm{k}^2  } .
\end{eqnarray}
It is important to keep in mind that $\CAtd$ is a function of the momentum $k$ carried by the gauge field.

We want to evaluate $\Pi_{(\rm A)}^{(b + c)}$:
\begin{eqnarray}
&&
\Pi^{(b+c)}_{(\text{A}) }(p)= \frac{p_\lambda  \epsilon_{\mu \nu \lambda}}{ {\color{black}2} |p|}
 \Pi^{(b+c)}_{\mu \nu} (p)   \nn \\
&&
=
\frac{p_\lambda  \epsilon_{\mu \nu \lambda}}{ {\color{black}2} |p|}
\times
2    \frac{(-1) g^4}{N_f}  
\int \frac{d^3k}{(2\pi)^3}
\int \frac{d^3q}{(2\pi)^3}
\Tr[ \gamma_\mu 
   \frac{ i\, q^\rho \gamma_\rho }{q^2}  \gamma_\alpha  
    \frac{ i\, (k+q)^\tau \gamma_\tau }{(k+q)^2}    \gamma_\beta
     \frac{ i\, q^\sigma \gamma_\sigma  }{q^2}   \gamma_\nu  
     \frac{ i\,( p+q)^{\kappa} \gamma_\kappa }{(p+q)^2}
   ] \,
D^{\text{Anti}}_{\alpha \beta }(k).  \qquad \;\;\; \cr
&&
\label{anti-bc-Diagram-Zero-Coulomb}
\end{eqnarray}
The following identity is useful:
\begin{eqnarray}
\epsilon_{ijk} \epsilon_{abc} 
=\text{Det}
\begin{pmatrix}
\delta_{i a} & \delta_{i b} & \delta_{i c}  \\ 
\delta_{j a} & \delta_{j b} & \delta_{j c}  \\ 
\delta_{k a} & \delta_{k b} & \delta_{k c}  \\ 
\end{pmatrix}.
\end{eqnarray}

Consider the following terms in the integrand in Eq.~\eqref{anti-bc-Diagram-Zero-Coulomb}:
\begin{eqnarray}
&&   
\Tr[ \gamma_\mu 
   \frac{ i\, q^\rho \gamma_\rho }{q^2}  \gamma_\alpha  
    \frac{ i\, (k+q)^\tau \gamma_\tau }{(k+q)^2}    \gamma_\beta
     \frac{ i\, q^\sigma \gamma_\sigma  }{q^2}   \gamma_\nu  
     \frac{ i\,( p+q)^{\kappa} \gamma_\kappa }{(p+q)^2}
   ] \,
\,  \, (  p_\lambda  \epsilon_{\mu \nu \lambda}  )
\, \frac{( \epsilon_{ \alpha \beta \eta}  \,   k_{\eta} \delta_{\eta j}  ) }{\bm{k}^2}  \\
&&
=
\frac{1}{ q^2 (k+q)^2  q^2  (p+q)^2 }
\Tr[\gamma_\mu \gamma_\rho \gamma_\alpha \gamma_\tau 
\gamma_\beta \gamma_\sigma \gamma_\nu \gamma_\kappa]
\,q^\rho (k+q)^\tau q^\sigma( p+q)^{\kappa}
\, (  p_\lambda  \epsilon_{\mu \nu \lambda}  )
\,  \frac{( \epsilon_{ \alpha \beta \eta}  \,   k_{\eta} \delta_{\eta j}  ) }{\bm{k}^2} \quad \quad \quad \\
&&
=
\frac{ 1 }{ q^2 (k+q)^2  q^2  (p+q)^2\;   }
\times
8 \, 
\Big(
[   q^2 (p \cdot q) + p^2 q^2 ] \, \bm{k}^2   
+ [ (p \cdot q) \, q^2 +  p^2 q^2  ]  \, \bm{k \cdot q}
\Big) 
 \frac{1}{\bm{k}^2}.
\label{tmp-bc-integral}
\end{eqnarray} 
We perform the convergent $q$-integral first:
\begin{eqnarray}
\int \frac{d^3q}{(2\pi)^3}   \Big(   \text{Eq}.~\eqref{tmp-bc-integral} \Big) 
\, && =
\int \frac{d^3q}{(2\pi)^3}  
\frac{ 1 }{   (k+q)^2  q^2  (p+q)^2\;  } \cr
&& \times
8 \, 
\Big(
[     (p \cdot q) + p^2   ] \, \bm{k}^2   
+ [ (p \cdot q) \,  +  p^2    ]  \, \bm{k \cdot q}
\Big) 
\frac{1}{\bm{k}^2}  .
\end{eqnarray}
Look at
\begin{eqnarray}
&&
\int \frac{d^3q}{(2\pi)^3}  
\frac{ p\cdot q + p^2 }{   (k+q)^2  q^2  (p+q)^2\;  }
=
\int \frac{d^3q}{(2\pi)^3}  
\frac{ \frac{1}{2} [(p+q)^2 -p^2-q^2 ]+ p^2   }{   (k+q)^2  q^2  (p+q)^2\;  }  \cr
&&
=
\frac{1}{2}
\int \frac{d^3q}{(2\pi)^3}  
\Big(
 \frac{1}{(k+q)^2  q^2    }
 - \frac{1}{  (k+q)^2     (p+q)^2\;   }
 +
 \frac{   p^2   }{   (k+q)^2  q^2  (p+q)^2\;  } 
\Big)  \cr
&&
=
\frac{1}{2}
(  \frac{1}{8|k|}  -  \frac{1}{8|p-k|}   + \frac{p^2}{ 8|p||k||p-k|}),
\end{eqnarray}
where we used a formula in Appendix \ref{loop-integral-formula}.
Next consider
\begin{eqnarray}
\int \frac{d^3q}{(2\pi)^3}  
\frac{ (p\cdot q + p^2 )  \, \bm{k \cdot q}  }{   (k+q)^2  q^2  (p+q)^2\;  }
\, && =
\frac{1}{2}
\int \frac{d^3q}{(2\pi)^3}  
\Big(
 \frac{ \bm{k \cdot q} }{(k+q)^2  q^2    }
 - \frac{ \bm{k \cdot q} }{  (k+q)^2     (p+q)^2\;   } \cr
 &&
 +
  \frac{   p^2 \bm{k \cdot q}  }{   (k+q)^2  q^2  (p+q)^2\;  } 
\Big)    \nn \\
&&
=
\frac{1}{2}
\Big(
\frac{- \bm{k}^2  }{16 |k|}
+
\frac{ \bm{k}^2 + \bm{k \cdot p} }{16  |k-p|}
+
p^2 \, 
L(\bm{k},p,k)
\Big)  
\qquad
\end{eqnarray}
where we used Eq.~\eqref{3-propagator-linear-q}.
Therefore,
\begin{eqnarray}
&&
\int \frac{d^3q}{(2\pi)^3}   \Big(  \text{ Eq}.~\eqref{tmp-bc-integral} \Big) \cr
&& =
\frac{8}{2}
(  \frac{1}{8|k|}  -  \frac{1}{8|p-k|} + \frac{p^2}{ 8|p||k||p-k|}  )
+
\frac{8}{\bm{k}^2} 
\frac{1}{2}
\Big(
\frac{- \bm{k}^2  }{16 |k|}
+
\frac{ \bm{k}^2 + \bm{k \cdot p} }{16  |k-p|}
+
p^2 \,  L(\bm{k},p,k)
\Big)     \quad
\\
&&
=
\frac{1}{2}
(  \frac{1}{ |k|}  -  \frac{1}{ |p-k|}  + \frac{p^2}{  |p||k||p-k|} )
+
\frac{1}{2}
\Big(
\frac{- 1  }{2 |k|}
+
\frac{ 1 + \frac{\bm{k \cdot p}}{\bm{k}^2}   }{2 |k-p|}
\Big) 
+
\frac{4 \, p^2 }{\bm{k}^2}  \,  L(\bm{k},p,k)    \\
&&
= \frac{1}{4}
(  \frac{1}{|k|}   +   \frac{ -1 + \frac{\bm{k \cdot p}}{\bm{k}^2}   }{  |k-p|}   )
+
 \frac{p^2}{ 2\, |p||k||p-k|}
 +
\frac{4 \, p^2 }{\bm{k}^2}  \,  L(\bm{k},p,k) ,
\end{eqnarray}
where we used
\begin{eqnarray}
&&
\int \frac{d^3q}{(2\pi)^3}    \frac{ \bm{k \cdot q} }{(k+q)^2  q^2    }  
=  \frac{ - \bm{k^2} \,  }{ 16 |k|} , \\
&&
\int \frac{d^3q}{(2\pi)^3}    \frac{ \bm{k \cdot q} }{ (q+k)^2 (q+p)^2    }
=
- \frac{ \bm{k}^2 + \bm{k \cdot p} }{16  |k-p|} .
\end{eqnarray}

Summarizing, we find that Eq.~\eqref{anti-bc-Diagram-Zero-Coulomb} equals
\begin{eqnarray}
\Pi^{(b+c)}_{(\text{A}) }(p)
\, && =
\frac{ 1 }{ {\color{black}2}   |p|}
\; {\color{black}2}    \frac{(-1) g^4}{N_f}  
\Big(
\int   \CAtd(k)   \frac{ dk_0 d^2\bm{k}}{(2\pi)^3} \;
\frac{1}{4}
(  \frac{1}{|k|}   +   \frac{ -1 + \frac{\bm{k \cdot p}}{\bm{k}^2}   }{  |k-p|}   )
+
 \frac{p^2}{ 2\, |p||k||p-k|} \cr
&& +
\frac{4 \, p^2 }{\bm{k}^2}  \,  L(\bm{k},p,k) 
\; \;
\Big)   \nn \\
&&
=
\frac{ 1 }{ {\color{black}2} |p|}
\; 2    \frac{(-1) g^4}{N_f}  
\Big( 
\int  \CAtd(k)   \frac{ dk_0 d^2\bm{k}}{(2\pi)^3} \;
\frac{1}{4}
\frac{ \bm{k \cdot p}   }{  |k-p|  \;\bm{k}^2 }
+
 \frac{p^2}{ 2\, |p||k||p-k|}
 +
\frac{4 \, p^2 }{\bm{k}^2}  \,  L(\bm{k},p,k) \cr
&& +
\frac{1}{4 |k|}
-
\frac{1}{4 |k-p|}
\Big).
\label{Pibc-q-out}
\end{eqnarray}

\subsection{Anti-symmetric component of $ \Pi^{(a)}$}

We aim to evaluate
\begin{eqnarray}
\Pi^{(a)}_{(\text{A}) }(p) && = \frac{p_\lambda  \epsilon_{\mu \nu \lambda}}{ {\color{black}2} |p|}
 \Pi^{(a)}_{\mu \nu} (p)   \nn \\
&&
=
  \frac{p_\lambda  \epsilon_{\mu \nu \lambda}}{ {\color{black}2}  |p|}
  \frac{(-1) g^4}{N_f}  
\int_k
\int_q
 \Tr[ \gamma_\mu G(q) \gamma_\alpha G(q+k) \gamma_\nu G(q+k+p) \gamma_\beta G(p+q)] \cr
 && \times
\, \frac{ \CAtd(k) \, \epsilon_{ \alpha \beta \lambda} k_{\eta} \delta_{\eta j}  }{\bm{k^2}}
 \\
&&
 =
  \frac{ 1 }{ {\color{black}2}  |p|}
  \frac{(-1) g^4}{N_f}  
\int_k \CAtd(k)
\int_q
\Big[
 -8 (k\cdot p )\,( p \cdot q) k_\eta - 8( p\cdot q)^2 k_\eta + 8 (k \cdot q) p^2 k_\eta   \nn \\
 &&
 +  4 (p \cdot q) p^2 k_\eta + 12 p^2 q^2 k_\eta + 8 (k \cdot q) (p \cdot q) p_\eta + 
 4 k^2( p\cdot q) p_\eta - 4 (k \cdot p) q^2 p_\eta - 4 (k \cdot p) p^2 q_\eta 
   \nn \\
&&   
 - 8 (k\cdot q) p^2 q_\eta - 4 k^2 p^2 q_\eta
\Big]  
\times  
\frac{1}{ q^2 (q+k)^2  (q+k+p)^2 (q+p)^2}
\times
\, \frac{    k_{\eta} \delta_{\eta j}  }{\bm{k^2}}   .
   \label{pia-before-dot-with-Dmunu}
 \end{eqnarray}
The integral over $q$ in Eq.~\eqref{pia-before-dot-with-Dmunu} can be decomposed as
\begin{eqnarray}
\mathcal{I_C}
\equiv 
\int_q
  \frac{ C_1 (p \cdot q) + C_2 (k \cdot q) + C_3 \,q^2 + C_4 (k \cdot q)(p \cdot q) 
 + C_5 \bm{(k \cdot q)} + C_6  (k \cdot q)  \bm{(k \cdot q) } 
 + C_7 \, (p \cdot q)^2  }
{ q^2 (q+k)^2  (q+k+p)^2 (q+p)^2}, \cr
\end{eqnarray}
for some $C_i$, independent of $q$.
Using the building block integrals from Appendix \ref{all-fomula-needed} and writing the norms $|k| = k$, and $|p| = p$, we find 
\begin{eqnarray}
\mathcal{I_C}
&& =
{\color{black}  
(  \frac{-1}{2  |k| } +   \frac{1}{4 |k-p|}  + \frac{1}{4 |k+p|}  )
}
+    ( \frac{2 p }{ 4 |k| \, |k-p|}  + \frac{3 p  }{4 |k| \, |k+p|}  )   
+
{\color{black}
\frac{- k^2 \,p}{ 2 |k| \, |k-p| \,(k \cdot p)}
} \nn \\
&& +
\frac{- p^3 }{ 2 |k| \, |k-p| \,(k \cdot p)}
+
{\color{black}
\frac{  k^2 \,p}{ 2 |k| \, |k+p| \,(k \cdot p)}
}
+
\frac{ p^3 }{ 2 |k| \, |k+p| \, (k \cdot p)}
+
\frac{ -p^2 }{ 8 |k| \, ( |p|\, |k| + k \cdot p ) }    \nn \\
&&
+
\frac{ p^2  }{  8 |k-p| \, ( |p|\, |k| + k \cdot p )  }
+
\frac{ p^3 }{  8 \,|k| \, |k-p| \, ( |p|\, |k| + k \cdot p )  } +
\frac{ -p  }{ 4 \big(   k^2 + |k| \, |k+p|  +  k \cdot p   \big)  }  \nn \\
&&
+
\frac{ p^2  }{ 8 k\,\big(   k^2 + |k| \, |k+p|  +  k \cdot p   \big)  } +
\frac{p^2 }{ 8 |k+p| \, \big(   k^2 + |k| \, |k+p|  +  k \cdot p   \big)     }
 \nn  \\
&&
+
\frac{ -|p| \, |k|  }{ 8 |k+p| \,  \big(   k^2 + |k| \, |k+p|  +  k \cdot p   \big)  } 
+
\frac{ -p \, |k+p|  }{ 8 |k|\,  \big(   k^2 + |k| \, |k+p|  +  k \cdot p   \big)   }  
  \nn  \\
&&   
+
\frac{ p^2  }{  8 |k| \, \big( p^2 +   |p| \, |k-p| - k \cdot p \big)  }
+
\frac{ -p^2 }{ 8 |k-p| \,  \big( p^2 +   |p| \, |k-p| - k \cdot p \big)   }
 \nn  \\
&&  
+
\frac{ p^3 }{ 8 |k| \, |k-p| \,\big( p^2 +   |p| \, |k-p| - k \cdot p \big)    } 
+
\frac{ p^2 }{  8 |k| \, \big(  p^2 + |p| \, |k+p| \, +  k \cdot p    \big)  } \cr
&&
+
\frac{ -p^2 }{  8 |k+p| \, \big(  p^2 + |p| \, |k+p| \, +  k \cdot p    \big)  }
+
\frac{1}{  8 |k| \, |k+p| \, \big( p^2 +|p| \, |k+p| +   k \cdot p     \big)}  .
\label{Ic-form}
\end{eqnarray}
Note that, throughout the calculation, we have kept $p$ general. 
Many terms above are individually divergent when integrated over $k$.
We cut off the divergent $k$ integrals with the cutoff $\Lambda$:
For instance,
\begin{eqnarray}
&&  \int_{-1}^1 d\cos\theta  \int_0^{\Lambda}  dk  \frac{  2\pi k^2  }{(2\pi)^3} \frac{1}{k} 
= \frac{ \Lambda^2 }{4 \pi^2} , \\
&&  \int  \frac{  d^3k  }{(2\pi)^3} 
 \frac{1}{ |k-p|} 
= \frac{ \Lambda^2 }{4 \pi^2} 
+ \frac{ -p^2 }{ 12 \pi^2}   ,
\label{Ld-sq-div}      \\
&&
{\color{black}
\int \frac{d^3k}{(2\pi)^3}
\frac{- k^2 \,p}{ 2 |k| \, |k-p| \,(k \cdot p)}
}
= \frac{p^2 (12+\pi^2)}{64 \pi^2}
+ 0 \cdot \log \Lambda
+ \frac{-p \Lambda}{4\pi^2}.
\end{eqnarray}
We are careful not to shift the momenta arbitrarily in any divergent integral; otherwise, we are liable to obtain an incorrect result.

\subsection{Combining the anti-symmetric components of $\Pi^{(a)}$ and $\Pi^{(b+c)}$}

We now add together the anti-symmetric components of $\Pi^{(a)}$ and $\Pi^{(b+c)}$ in \eqref{pia-before-dot-with-Dmunu} and \eqref{Pibc-q-out} to find:
\begin{eqnarray}
&&
\Pi^{(a+b+c)}_{(\text{A}) }(p)
=
- \frac{ 1 }{ {\color{black}2}  |p|}
  \frac{ g^4}{N_f} \int_k  \nn \\
&&
\times   
 \CAtd(k)  \Big[
  2 \times \big(  \frac{1}{4} \frac{ \bm{k \cdot p}   }{  |k-p|  \;\bm{k}^2 }
+   \frac{p^2}{ 2\, |p||k||p-k|}
 +  \frac{4 \, p^2 }{\bm{k}^2}  \,  L(\bm{k},p,k) 
 + \frac{1}{4 |k|}
-
\frac{1}{4 |k-p|} 
  \; \big)
+
\mathcal{I_C} 
\;    \Big]  .  \,
\qquad    \;\; 
  \label{Pi-b-c-a-Anti}
\end{eqnarray}
To perform the integral over $k$, we take $p$ to lie along the $k_\tau$ axis, with $k \cdot p = |k||p| \cos\theta$, so that ${\bf k} \cdot {\bf p} = 0$ and
\begin{eqnarray}
 \int \frac{d^3k}{(2\pi)^3} (\ldots)
 =  \int_{-1}^{1 } d\cos\theta  \int_{0}^{\Lambda} dk\frac{ 2\pi k^2   }{(2\pi)^3}
 (\dots).
\end{eqnarray}
After performing the $\int_{0}^{\Lambda} dk$ integral, we perform a $\frac{1}{\Lambda}$ expansion, and then do the $\int_{-1}^{1 } d\cos\theta $ integral. The $\log \Lambda$ divergent terms vanish after the angular integration.
Letting $\cos\theta =z$, we symmetrize with respect to $z$ to get rid of terms that are odd in $z$ and should therefore vanish after performing the $z$ integral.
The result is
\begin{eqnarray}
\Pi^{(a+b+c)}_{(\text{A}) }(p)
\, && =
 \frac{ 1 }{ {\color{black}2}  |p|}
  \frac{(-1) g^4}{N_f}     
  \int_{-1}^{1} dz
\, 
|p|^2 \,
\frac{ -z + (-1 + z^2) \text{ArcTanh}[z]    }{   8 \pi^2 \,z} \cr
 && \times
 \frac{1}{ A_X + B_X \sqrt{1-z^2} } ,   \\
 &&
 \sigma^\psi_{xy} = 
 \frac{i}{\omega} \Pi^{(b+c+a)} (\bm{p} = 0,|p_0|\to i |\omega|),
\end{eqnarray}
with $A_X \equiv  \frac{ \kappa^2+ \gX^4 }{ \kappa    }   ,
B_X   \equiv \frac{ \wx \gX^2  }{\kappa}   $.

As a consistency check: When $\wx =0$ (vanishing Coulomb interaction), 
\begin{eqnarray}
&&
\Pi^{(a+b+c)}_{(\text{A}) }(p; \wx =0)
=  \frac{ 1 }{ {\color{black}2}  |p|}
  \frac{(-1) g^4}{N_f}       \frac{ \kappa   }{\kappa^2+ \gX^4     }
 \;(   -p^2  \; \frac{4+ \pi^2}{32\pi^2}  
+ 0 \cdot \log \Lambda
+ \frac{p }{4\pi^2} \, \Lambda  
  \;  )    ,
\end{eqnarray}
which agrees with \cite{SPIRIDONOV1991109}.
We drop the linear divergence since it is an artifact of the (gauge-noninvariant) hard cutoff.

\subsection{Symmetric component of $\Pi^{(b+c)}$}

Now we consider the symmetric component of $\Pi^{(b+c)}$.
We will need the symmetric part of the gauge field propagator:
 \begin{eqnarray}
 &&  \label{D00-Sym}
 D^\text{Sym}_{00}(k_0,\bm{k})
=
\frac{ |k| }{ \bm{k}^2  }
\times  
\frac{1}{ \frac{ \kappa^2+ \gX^4  }{ \gX^2  } 
  + \wx \, \sqrt{1-\cos^2\theta}   }
(  1+  \, \frac{ \wx  }{ \gX^2}  \sqrt{1-\cos^2\theta}   )  ,  \\
&&  \label{Dij-Sym}
D^{\text{Sym}}_{ij}(k_0,\bm{k})
=
 \frac{ 1   }{  |k|   } \;
(\delta_{ij}- \frac{k_i k_j }{\bm{k^2}})
\times
\frac{1}{ \frac{ \kappa^2+ \gX^4   }{ \gX^2  } 
  + \wx   \sqrt{1-\cos^2\theta}     }.
 \end{eqnarray}
Here we are denoting $\sqrt{1 - \cos^2 \theta} = |{\bf k}|/|k|$.
We parameterize the gauge field propagator as 
\begin{eqnarray}
{\color{black} D^{\text{Sym}}_{\alpha \beta } (k)  }
=    \Big( \delta_{\alpha 0 } \delta_{\beta 0 } \, f_{A}(k)
 + \delta_{\alpha i } \delta_{\beta j }  \;\delta_{ij} f_B(k)  
 +    \delta_{\alpha i } \delta_{\beta j }  \; k_i k_j f_C(k) 
 \Big)  ,\;\;\;
\label{DSym-parameters}
\end{eqnarray}
where $f_A, f_B, f_C$ are constants.
Using the gauge field propagator above, the symmetric component of $\Pi^{(b+c)}$ is
\begin{eqnarray}
\Pi^{(b+c)}_{(\text{S}) }(p) \, &&= 
\frac{1}{2} \delta_{\mu \nu}  \Pi^{(b+c)}_{\mu \nu} (p)      \\
&&
=
\frac{ \delta_{\mu \nu} }{ 2  }
\times
2    \frac{(-1) g^4}{N_f}  
\int \frac{d^3k}{(2\pi)^3}
\int \frac{d^3q}{(2\pi)^3}
\Tr[ \gamma_\mu 
   \frac{ i\, q^\rho \gamma_\rho }{q^2}  \gamma_\alpha  
    \frac{ i\, (k+q)^\tau \gamma_\tau }{(k+q)^2}    \gamma_\beta \cr
    && \times
     \frac{ i\, q^\sigma \gamma_\sigma  }{q^2}   \gamma_\nu  
     \frac{ i\,( p+q)^{\kappa} \gamma_\kappa }{(p+q)^2}
   ] \,
 D^{\text{Sym}}_{\alpha \beta } (k)
 \qquad \;\;
\end{eqnarray}
The trace $\Tr[\ldots]_{\mu \nu}$ in the integrand evaluates to
\begin{eqnarray}
&&   
\Tr[ \gamma_\mu 
   \frac{ i\, q^\rho \gamma_\rho }{q^2}  \gamma_\alpha  
    \frac{ i\, (k+q)^\tau \gamma_\tau }{(k+q)^2}    \gamma_\beta
     \frac{ i\, q^\sigma \gamma_\sigma  }{q^2}   \gamma_\nu  
     \frac{ i\,( p+q)^{\kappa} \gamma_\kappa }{(p+q)^2}
   ] \,
\,  \, (  \frac{ \delta_{\mu \nu} }{ 2  } )
 D^{  \text{Sym}  }_{\alpha \beta }(k)   \label{Tr-Dlt-Dafbt}
  \\
&&
=
\frac{1}{q^2 (k+q)^2 q^2 (p+q)^2}
\Big(
\delta_{\alpha \beta}  
[  2 (k\cdot q) \, (p \cdot q)  - (k \cdot p) \, q^2 + (k \cdot q) \, q^2 
+ (p \cdot q) \, q^2  + q^4   \,]     \nn \\
&&
+ 
(   p_\alpha k_\beta + p_\beta k_\alpha   ) \, q^2
+
(  q_\alpha k_\beta + q_\beta k_\alpha  )  \, [ -2 (p\cdot q )  -q^2 ] +
 ( q_\alpha p_\beta  +    q_\beta p_\alpha   ) \, q^2 \cr
&&
 +
  (q_\alpha q_\beta)   \,  [-4 (p \cdot q)  - 2 q^2] 
\Big) \,
 D^{  \text{Sym}  }_{\alpha \beta }(k)      \nn    \\
&&
=
 \frac{1}{q^2 (k+q)^2 q^2 (p+q)^2}
\Big(
\delta_{\alpha \beta}  
[  2 (k\cdot q) \, (p \cdot q)  - (k \cdot p) \, q^2 + (k \cdot q) \, q^2 
+ (p \cdot q) \, q^2  + q^4   \,]       \nn \\
&&
+ 
(   p_\alpha k_\beta + p_\beta k_\alpha   ) \, q^2
+
(  q_\alpha k_\beta + q_\beta k_\alpha  )  \, [ -2 (p\cdot q )  -q^2 ]
+
 ( q_\alpha p_\beta  +    q_\beta p_\alpha   ) \, q^2 \cr
&&  +
  (q_\alpha q_\beta)   \,  [-4 (p \cdot q)  - 2 q^2] 
\Big) \,
{\color{black} D^{\text{Sym}}_{\alpha \beta } (k)  } . \nn \\
&&  
\end{eqnarray}
Next, we decompose the terms in the integrand with different $q$ dependencies into various partial fractions: 
\begin{eqnarray}
&&
\Tr[\ldots ]_{\mu \nu}  (  \frac{ \delta_{\mu \nu} }{ 2  } ) D^{\text{Sym}}_{\alpha \beta } (k)   \\
&&
=
\delta_{\alpha \beta} D^{\text{Sym}}_{\alpha \beta } (k) 
\Big[
\frac{1}{ 2 (q+k)^2 (q+p)^2 }
+ \frac{1}{2q^2 \,q^2}
+ \frac{-k^2}{2(q+k)^2 q^4 }
+ \frac{ -p^2}{ 2(q+p)^2  q^4}   \nn \\
&&
+ \frac{ k^2 \,p^2}{ 2(q+k)^2 (q+p)^2 \, q^4}
+ \frac{ - k\cdot p}{ (q+k)^2 (q+p)^2 \, q^2}
\Big]   \nn \\
&&
+
(   p_\alpha k_\beta + p_\beta k_\alpha   )     D^{\text{Sym}}_{\alpha \beta } (k) 
\frac{1}{ q^2 (q+k)^2 (q+p)^2}     \nn \\
&&
+
\Big(
\frac{ 
-2 f_B  (\bm{ k \cdot q }) -2 f_C \, \bm{k}^2  (\bm{ k \cdot q })  
        -2 f_A \, k_0 \, q_0  }{ (q+k)^2 q^4}
        +
\frac{ 2f_B \, p^2  (\bm{ k \cdot q } )  + 2 f_C\, p^2   \bm{k}^2  (\bm{ k \cdot q })  
       + 2 f_A  \, p^2  \, k_0 \,q_0    }{ (q+k)^2  (q+p)^2 \, q^4}        
\Big)  \nn \\
&&
+
\frac{ 2 f_B  (\bm{ p \cdot q } ) +2 f_A \, p_0 q_0  
 +  2 f_C \, (\bm{ k \cdot p } ) \, (\bm{ k \cdot q } ) 
}{ (q+k)^2 \, (q+p)^2 \, q^2}     \nn \\
&&
+
(-2) \frac{  f_B \bm{q}^2  + f_C \, (\bm{ k \cdot q } )^2 
          +    f_A \, q_0^2   }{ (q+k)^2 \, q^4 }
+ 
(+2)  \frac{ f_B \,p^2  \bm{q}^2  + f_C \, p^2 \,(\bm{ k \cdot q } )^2 
              +  f_A \, p^2 \, q_0^2  }{ (q+k)^2 (q+p)^2 \, q^4}.
\end{eqnarray}
Each of the terms in the first few lines diverge as $|q|\to 0^+$ in the IR, however, their sum is IR finite.
Combining some of these terms with one another, we perform the following $q$ integral:
\begin{eqnarray}
&&  
\int \frac{d^3q}{(2\pi)^3}
{\color{black} \frac{1}{2} }
\Big(
  \frac{1}{q^2 \,q^2}
+ \frac{-k^2}{(q+k)^2 q^4 }
+ \frac{ -p^2}{ (q+p)^2  q^4}
+ \frac{ k^2 \,p^2}{ (q+k)^2 (q+p)^2 \, q^4}  \Big) \\
&&
={\color{black} \frac{1}{2} }
\Big[
\int \frac{d^3q}{(2\pi)^3}
(  \frac{1}{q^2 \,q^2} + \frac{ -p^2}{ (q+p)^2  q^4})
+
\int \frac{d^3q}{(2\pi)^3}
(\frac{-k^2}{(q+k)^2 q^4 } + \frac{ k^2 \,p^2}{ (q+k)^2 (q+p)^2 \, q^4} )   \Big] \\
&&
={\color{black} \frac{1}{2} }
\Big[
\int \frac{d^3q}{(2\pi)^3}
\frac{ q^2 + 2 (q \cdot p)  }{ (q+p)^2  q^4 }
+
\int \frac{d^3q}{(2\pi)^3}
\frac{  -k^2 q^2 - 2 k^2 ( q \cdot p) }{ (q+k)^2 (q+p)^2 \, q^4 }  \Big]   \\
&&
={\color{black} \frac{1}{2} }
\Big[
\frac{1}{8 |p| }  -   \frac{ p \cdot p}{8 p^3}
+
\frac{-k^2 }{8} \frac{1}{ |p|  |k| |p-k|}
+(-2k^2) (-1) \frac{p^2 (p \cdot k)   + k^2 (p\cdot p) }{16 k^3 p^3 |k-p|}
\Big] \\
&& ={\color{black} \frac{1}{2} }
\frac{ p \cdot k  }{ 8 |k| \, |p| \, |k-p| },
\qquad  \;
\end{eqnarray}
where we have used the ``dot $p_5$" formulas in Appendix \ref{loop-integral-formula-part2}, with $p_5 =p$
The remaining integrals over $q$ are straightforwardly performed using formulas we have already given.

Next, we recall the following list of SO(3) non-invariant integrals that we evaluated in Appendix \ref{loop-integral-formula-part2}:
\begin{eqnarray}
&&  
\int \frac{d^3q}{(2\pi)^3}
\frac{ \bm{ k \cdot q } }{ (q+k)^2 q^4}
=
R_1 ( \bm{k}; k)     , \;\;\;\; p_5 = (0,\bm{k})  ,    \\
&&
\int \frac{d^3q}{(2\pi)^3}  \frac{ k_0 \, q_0  }{ (q+k)^2 q^4}
= R_1 ( k_0; k)    , \;\;\;\; p_5 = (k_0,0,0) ,
\end{eqnarray}
\begin{eqnarray}
&&
\int \frac{d^3q}{(2\pi)^3}
\frac{  (\bm{ k \cdot q } )  }{ (q+k)^2  (q+p)^2 \, q^4} 
 = N( \bm{k} ; p,k )  
       , 
\int \frac{d^3q}{(2\pi)^3}
\frac{ k_0 \,q_0    }{ (q+k)^2  (q+p)^2 \, q^4}   
=  N( k_0 ; p,k )   ,   \qquad  \;\;\;   
\end{eqnarray}
\begin{eqnarray}
&&
\int \frac{d^3q}{(2\pi)^3}
\frac{ (\bm{ p \cdot q } )  }{ (q+k)^2 \, (q+p)^2 \, q^2}  
= L(\bm{p};, k,p )
,   
\int \frac{d^3q}{(2\pi)^3}
\frac{  p_0 q_0   }{ (q+k)^2 \, (q+p)^2 \, q^2}  
=L(p_0;, k,p )   , \qquad  \;\;\; 
\end{eqnarray}
and
\begin{eqnarray}
&&     
\int \frac{d^3q}{(2\pi)^3}     
\frac{  (\bm{ k \cdot q } )^2    }{ (q+k)^2 \, q^4 }  
=
R_2 (\bm{k},\bm{k}; k) , \\ 
&& 
\int \frac{d^3q}{(2\pi)^3}
\frac{ \bm{q}^2  }{ (q+k)^2 \, q^4 } \cr
&& =\int \frac{d^3q}{(2\pi)^3}
\frac{ (q_x + iq_y) (q_x - iq_y)  }{ (q+k)^2 \, q^4 } 
=  \int \frac{d^3q}{(2\pi)^3}
\frac{ \Big( q \cdot (0,1,i) \Big) \, \Big( q \cdot (0,1,-i) \Big)  }{ (q+k)^2 \, q^4 }  \cr
&&
=
R_2 ((0,1,i) , (0,1,-i); k) 
= \frac{ 2 k^2 + \bm{k}^2   }{32 \,k^3},
\\
&&
\int \frac{d^3q}{(2\pi)^3}
\frac{   q_0^2 }{ (q+k)^2 \, q^4 }  
=
   \int \frac{d^3q}{(2\pi)^3}
\frac{ \big(   q \cdot (1,0,0) \; \big)^2 }{ (q+k)^2 \, q^4 }  \cr
&& =
R_2 ( (1,0,0) , (1,0,0); k)   
= \frac{  k^2 + k_0^2 }{32 k^3} .
\end{eqnarray}
Finally, we'll need:
\begin{eqnarray}
&&
 \int \frac{d^3q}{(2\pi)^3}
 \frac{ (\bm{ k \cdot q } )^2 }{ (q+k)^2 (q+p)^2 \, q^4}  
 =R_3 ( \bm{k} ; k,p) ,   \\
&&
 \int \frac{d^3q}{(2\pi)^3}
 \frac{ q_0^2  }{ (q+k)^2 (q+p)^2 \, q^4}
 =R_3 ( (1,0,0) ; k,p)  ,
\\
&&
\int \frac{d^3q}{(2\pi)^3}
 \frac{ \bm{q}^2  }{ (q+k)^2 (q+p)^2 \, q^4} \cr
\, && = 
 \int \frac{d^3q}{(2\pi)^3}
 \frac{  q^2 -q_0^2  }{ (q+k)^2 (q+p)^2 \, q^4} 
 =
  \frac{1}{8 |k|\, |p| \, |k-p|}  - R_3 ( (1,0,0) ; k,p)    .
  \qquad \;\; 
\end{eqnarray}

Plugging these in, we find the $q$ integral equals
\begin{eqnarray}
&&  
I_{bc}
\equiv
\int \frac{d^3q}{(2\pi)^3}
\Tr[\ldots]_{\mu \nu}  (  \frac{ \delta_{\mu \nu} }{ 2  } )
 D^{\text{Sym}}_{\alpha \beta } (k)   \nn \\
&&
=
(f_A + 2 f_B + f_C  \,\bm{k^2} )
[\frac{J_2(p,k)}{2}  +  (- p \cdot k)\frac{J_3(p,k) }{2}  ]
+
[2f_B \bm{k \cdot p} + 2f_C\bm{k^2} \bm{k \cdot p} + 2f_A k_0 p_0 ] J_3(p,k)  \nn  \\
&&
+
\frac{f_A \,k_0^2+ f_B \bm{k}^2  + f_C \bm{k^4}   }{ 8 k^3}
+
\big[   2 f_B p^2 N(\bm{k}; p,k) + 2 f_C \bm{k^2} p^2  \, N(\bm{k}; p,k)
+ 2 f_A \, p^2  \, N(k_0; p,k)  \; \big]   \nn  \\
&&
+ 
\big[
 2 f_B \, L(\bm{p}; k,p) + 2f_A \, L(p_0; k, p )
 +   2 f_C (\bm{k \cdot p})  \, L(\bm{k}; k,p)
\big]   \nn \\
&&
+
(-1)
\big[
f_B  \frac{ 2k^2+ \bm{k^2}}{16 k^3}
+ f_C \bm{k^2}   \frac{  k^2+ \bm{k^2}}{16 k^3}
+ f_A    \frac{  k^2+  k_0^2}{16 k^3}
\big]   \nn \\
&&
+
2 p^2 \,
\Big(
f_B \,  \, [ J_3(k,p)- R_3 ( (1,0,0) ; k,p)   \; ]
+ f_C    \, R_3 ( \bm{k} ; k,p) 
+ f_A   \, R_3 ( (1,0,0) ; k,p) 
\Big)
\label{Pi-bc-Sym}
\end{eqnarray}
Note that in the definition of $I_{bc}$ above, we do not include the factor $2$ which counts the contribution from diagrams $b$ and $c$.

\subsection{Symmetric component of $\Pi^{(a)}$}

The symmetric component of $\Pi^{(a)}$ is
\begin{eqnarray}
&&  
\Pi^{(a)}_{(\text{S}) }(p)= 
\frac{1}{2} \delta_{\mu \nu}  \Pi^{(a)}_{\mu \nu} (p)      \cr
&&
=
\frac{ \delta_{\mu \nu} }{ 2  }
\times
\frac{(-1) g^4}{N_f}  
\int \frac{d^3k}{(2\pi)^3}
\int \frac{d^3q}{(2\pi)^3}
 \Tr[ \gamma_\mu G(q) \gamma_\alpha G(q+k) \gamma_\nu G(q+k+p) \gamma_\beta G(p+q)]
\, D^{  \text{Sym}  }_{\alpha \beta }(k)  \cr
&&
=
\frac{(-1) g^4}{N_f}  
\int \frac{d^3k}{(2\pi)^3}
\int \frac{d^3q}{(2\pi)^3}   \nn \\
&&   
\Bigg(
\delta_{\alpha \beta}
\big[ -4 (k \cdot p)( k \cdot q) -4 (k \cdot q)^2 -2 (k \cdot q ) (p \cdot q)+ 3 k^2 (p \cdot q)   
  - p^2 \,(k \cdot q)  -q^2 ( k \cdot p)    
      \nn \\
&&   
- 2 q^2 (k\cdot q) 
   + 3 k^2 \,q^2 \, -2 q^2 (p \cdot q) - p^2 \, q^2 - q^4    \, \big]
+ k_\alpha k_\beta [ -4 q^2 -4 (p \cdot q)  ]   \nn \\
&&
 + q_\alpha q_\beta [2 p^2-4 k^2] 
 + p_\alpha p_\beta  [2 q^2+ 2 (k\cdot q)]    \nn \\
&&
+
p_\alpha q_\beta [2 (k \cdot q) - 3 k^2+ 2q^2]
+ p_\beta q_\alpha  [ -2 (k \cdot p)  - 2 (k\cdot q) -k^2 -4 (p \cdot q) -2 q^2 ]  \nn \\
&&
+ k_\alpha q_\beta [4 (k\cdot p) + 4 (k\cdot q)  +p^2]
+ q_\alpha k_\beta [  4 (k\cdot q)  +p^2]
+ p_\alpha k_\beta [q^2 + 4 (k\cdot q)]
+ k_\alpha p_\beta [-2 (p \cdot q) -q^2 ]
\Bigg)  \qquad \;\; 
 \nn  \\
&&
\times 
\frac{1}{ q^2 (q+k)^2  (q+k+p)^2 (q+p)^2}
\, D^{  \text{Sym}  }_{\alpha \beta }(k).
\end{eqnarray}
Above, we have used the parameterization of the symmetric part of the gauge field propagator in Eq.~\eqref{DSym-parameters}.

We rewrite the first part of the integrand using partial fractions:
\begin{eqnarray}
&&
D^{ \rm sym}_{\alpha \beta } (k)
\delta_{\alpha \beta}
\Big(
\frac{2 (k \cdot p) \,k^2 -k^4 - \frac{3}{2} k^2 p^2 }{ q^2 (q+k)^2  (q+k+p)^2 (q+p)^2}
+ 
\frac{ 2k^2 }{ q^2 \, (q+k+p)^2 (q+k)^2} \cr
&& +
\frac{ {\color{black}-} 2 (k \cdot p) + 2k^2 - (q+k)^2 }{ q^2 \, (q+k+p)^2 (q+p)^2} 
+
\frac{ \frac{-1}{2} }{ (q+k+p)^2 (q+k)^2}
+
\frac{ \frac{ 3}{2} }{ (q+k+p)^2 (q+p)^2}
+
\frac{ \frac{-1}{2} }{ (q+k+p)^2 q^2} \cr
&& +
\frac{  (k \cdot p)  - \frac{q^2}{2} }{ (q+k+p)^2 (q+k)^2 (q+p)^2 }
\Big) .
\end{eqnarray}
Integrating this over $q$ gives:
\begin{eqnarray}
&&
D^{ \rm sym}_{\alpha \beta } (k)
\delta_{\alpha \beta}
\Big(
[2 (k \cdot p) \,k^2 -k^4 - \frac{3}{2} k^2 p^2] \,J_4(k,p)
+ 2 k^2 \,  J_3(k+p,k) \cr
&&
+ [ {\color{black}-} 2 (k \cdot p) + 2k^2 ] \,  J_3(k+p,p)  
- I(p,-k)  
+  \frac{-1}{2} \, J_2(k+p,k)
+ \frac{3}{2}  \, J_2(k+p,p)
+ \frac{-1}{2} \, J_2(k+p,0) \cr
&&
+ (k \cdot p) \, J_3( k, k-p) 
+ \frac{-1}{2} \, {\color{black} I( k, p)}
\Big)  \\
&&
=
D^{ \rm sym}_{\alpha \beta } (k)
\delta_{\alpha \beta}
\Big(
[  -k^4 - \frac{3}{2} k^2 p^2] \,J_4(k,p)
+
\frac{ |k| }{4 p |k-p|}  -  \frac{|k|  }{4 p |k+p|} 
+
\frac{-1}{4 p}  + \frac{1}{16 |k-p|} + \cr
&&  \frac{1}{16 |k+p|}
 +  \frac{ |k| }{2 p  \, |k+p|}
\Big)  \\
&&
= 
D^{ \rm sym}_{\alpha \beta } (k)
\delta_{\alpha \beta}
\Big(
[  -k^4 - \frac{3}{2} k^2 p^2] \,J_4(k,p)
+
0
+
\frac{-1}{4 p}  + \frac{1}{16 |k-p|} +  \frac{1}{16 |k+p|}
 +  \frac{ |k| }{2 p  \, |k+p|}
\Big) .
\end{eqnarray}

Next, we use partial fractions to re-express:
\begin{eqnarray}
&&
D^{ \rm sym}_{\alpha \beta } (k) 
 k_\alpha k_\beta   
 \frac{ -4 q^2 -4 (p \cdot q) }{q^2 (q+k)^2  (q+k+p)^2 (q+p)^2  }  \nn \\
&& 
 =
(D^{ \rm sym}_{\alpha \beta } (k)    k_\alpha k_\beta   ) 
\Big(
\frac{2 p^2 }{q^2 (q+k)^2  (q+k+p)^2 (q+p)^2 } 
-  \frac{2}{  (q+k)^2  (q+k+p)^2 (q+p)^2 } \cr
&&
- \frac{2}{  (q+k)^2  (q+k+p)^2 q^2 }
\Big)   .
\end{eqnarray}
Integrating this over $q$ gives:
\begin{eqnarray}
&&
(D^{ \rm sym}_{\alpha \beta } (k)    k_\alpha k_\beta   ) 
\Big(
2 p^2 \, J_4 (k,p)
-2 \, J_3 (p,p-k)
-2  \, J_3 (k,k+p)
\Big)    \cr
&&
=
(D^{ \rm sym}_{\alpha \beta } (k)    k_\alpha k_\beta   ) 
\Big(
\frac{ p }{4 |k| \, |k-p| (k \cdot p)} 
+ \frac{- p }{4 |k| \, |k+p| (k \cdot p)}
+  \frac{ -1 }{ 4|k|\, |p| \, |k-p|} \cr
&&
+   \frac{ -1 }{ 4|k|\, |p| \, |k+p|}
\Big).
\end{eqnarray}

The next term is
\begin{eqnarray}
&& D^{ \rm sym}_{\alpha \beta } (k) 
q_\alpha q_\beta 
 \frac{ ( 2 p^2-4 k^2 ) }{q^2 (q+k)^2  (q+k+p)^2 (q+p)^2  } \cr
 &&
=
(2 p^2-4 k^2)
\frac{ f_C (\bm{k \cdot q})^2 + f_A q_0^2 + f_B \, \bm{q}^2     }
{ q^2 (q+k)^2  (q+k+p)^2 (q+p)^2   } \cr
&& =
(2 p^2-4 k^2)
\frac{ f_C (\bm{k \cdot q})^2 + (f_A -f_B) q_0^2 + f_B \, q^2     }
{ q^2 (q+k)^2  (q+k+p)^2 (q+p)^2   }  .
\end{eqnarray}
Integrating this over $q$ gives:
\begin{eqnarray}
&&
(2 p^2-4 k^2)
\Big(\, 
f_C \, M_2(\bm{k};k,p)
+
(f_A- f_B)\, M_2( (1,0,0);k,p)
+ f_B \, J_3(k-p,k)
\Big) .
\end{eqnarray}

Moving on to the next term:
\begin{eqnarray}
&&
 D^{ \rm sym}_{\alpha \beta } (k) 
p_\alpha p_\beta   
\frac{ 2 q^2+ 2 (k\cdot q)  }{ q^2 (q+k)^2  (q+k+p)^2 (q+p)^2    }   \nn \\
&&
=
 D^{ \rm sym}_{\alpha \beta } (k) 
p_\alpha p_\beta  
\Big(
\frac{ -k^2 }{ q^2 (q+k)^2  (q+k+p)^2 (q+p)^2   }
+
\frac{1}{ (q+k+p)^2 (q+k)^2   (q+p)^2 } \cr
&& +
\frac{1}{ (q+k+p)^2 (q+p)^2  \, q^2 }
\Big) .
\end{eqnarray}
Integrating over $q$ gives
\begin{eqnarray}
&&
 D^{ \rm sym}_{\alpha \beta } (k) 
p_\alpha p_\beta  
\Big( \,
-k^2 \, J_4 (k,p)
+ J_3 (k,k-p)
+ J_3 (p,k+p)
\Big)    \\
&&
=
[f_C \bm{(k \cdot p)}^2 + f_A \, p_0^2  + f_B \, \bm{p^2}  ]
\;
\Big( \,
\frac{ -|k| }{ 8 |p| \, |k-p| \, (k \cdot p)}
+ \frac{ |k| }{ 8 |p| \, |k+p| \, (k \cdot p)}
+
\frac{1}{ 8 |k| \, |p| \, |k-p|} \cr
&&
+ \frac{1}{ 8 |k| \, |p| \, |k+p|}
\Big) .
\end{eqnarray}

Next, we consider (arranging the terms by their power of $q$):
\begin{eqnarray}
&&  
 D^{ \rm sym}_{\alpha \beta } (k) 
 \frac{ p_\alpha q_\beta [\ldots] +  q_\alpha p_\beta [\ldots]    }{  q^2 (q+k)^2  (q+k+p)^2 (q+p)^2  } 
   \nn \\
&&   
 =
\frac{ -4 f_C (\bm{k \cdot p}) \, (p \cdot q) \, (\bm{k \cdot q})
   - 4 f_B \, (p \cdot q)   \, (\bm{p \cdot q})  -4 f_A\, (p \cdot q) \,(p_0 q_0)   }
{  q^2 (q+k)^2  (q+k+p)^2 (q+p)^2 }     \cr
&&
+
\frac{ [-2 f_B (k \cdot p) - 4 f_B \, k^2]   (\bm{p \cdot q} )  }{q^2 (q+k)^2  (q+k+p)^2 (q+p)^2 } \nn \\
&& +
\frac{[-2 f_C (k \cdot p ) (\bm{k \cdot p} ) - 4 f_C\, k^2 \, (\bm{k \cdot p} )] \,
  (\bm{k \cdot q} )   }{q^2 (q+k)^2  (q+k+p)^2 (q+p)^2} 
+
\frac{ [-2 f_A (k \cdot p ) - 4 f_A k^2   ]( p_0\,q_0) }{q^2 (q+k)^2  (q+k+p)^2 (q+p)^2} .
\end{eqnarray}
Integrating over $q$ gives
\begin{eqnarray}
&&
   -4 f_C (\bm{k \cdot p}) \,F_1  -4 f_B \,F_2 -4 f_A \, F_3  
+
[-2 f_B (k \cdot p) - 4 f_B \, k^2] \, M(p;k )   \nn\\
&&
+
[-2 f_C (k \cdot p ) (\bm{k \cdot p} ) - 4 f_C\, k^2 \, (\bm{k \cdot p} )] \, M(k; p)
   \nn \\
   &&
+
[-2 f_A (k \cdot p ) - 4 f_A k^2   ]
\big[ \frac{1}{2}J_3(k,k+p)
  - \frac{1}{2}  J_3 (k-p,k)
  -\frac{p^2 }{2}  J_4(k,p)
  - M(p;k)
 \big],
\end{eqnarray}
where $F_1$, $F_2$, and $F_3$ are defined below:
\begin{eqnarray}
F_1 \, && \equiv
 \int_q
\frac{  (p \cdot q) \, (\bm{k \cdot q})  }{  q^2 (q+k)^2  (q+k+p)^2 (q+p)^2 } \cr
&&
=
 \int_q
  \frac{1}{2} 
\frac{ [(q+p)^2 -p^2 -q^2 ] (\bm{k \cdot q})  }{  q^2 (q+k)^2  (q+k+p)^2 (q+p)^2 }   \cr
&&
=
  \frac{1}{2} 
  \int_q
  \Big(
  \frac{   (\bm{k \cdot q})  }{  q^2 (q+k)^2  (q+k+p)^2   } 
  - \frac{ p^2 (\bm{k \cdot q})  }{  q^2 (q+k)^2  (q+k+p)^2 (q+p)^2 } \cr
  &&
  -\frac{   (\bm{k \cdot q})  }{    (q+k)^2  (q+k+p)^2 (q+p)^2 }
  \Big)   \cr
&&  
  =
  \frac{1}{2}
  \Big(
  L(\bm{k}; k,k+p) - p^2 \,M(k;p)  
  -L(\bm{k}; p,p-k)
  + \bm{k}^2 \,J_3 (p,p-k)
  \Big) ,
\end{eqnarray}
\begin{eqnarray}
F_2 \, && \equiv
\int_q \frac{ (p \cdot q)   \, (\bm{p \cdot q})   }{  q^2 (q+k)^2  (q+k+p)^2 (q+p)^2 }    \cr
&& =
\frac{1}{2}
\int_q \frac{ [ (q+p)^2  -p^2 -q^2 ]   \, (\bm{p \cdot q})   }{  q^2 (q+k)^2  (q+k+p)^2 (q+p)^2 }   
\cr
&&
=
\frac{1}{2}
\Big(
\int_q \frac{    \, (\bm{p \cdot q})   }{  q^2 (q+k)^2  (q+k+p)^2  }   
-
\int_q \frac{  p^2  \, (\bm{p \cdot q})   }{  q^2 (q+k)^2  (q+k+p)^2 (q+p)^2 }  
- \cr
&&
\int_q \frac{     \, (\bm{p \cdot q})   }{   (q+k)^2  (q+k+p)^2 (q+p)^2 }
\Big)  \cr
&&
=
\frac{1}{2}
\Big(
L(\bm{p}; k,k+p) -p^2 \,M(p;k)
-L(\bm{p}; k,k-p )
+ \bm{p}^2 \, J_3 (k,k-p)
\Big),
\end{eqnarray}
\begin{eqnarray}
F_3 \, && \equiv
\int_q
\frac{ (p \cdot q) \,(p_0 q_0)   }{  q^2 (q+k)^2  (q+k+p)^2 (q+p)^2 }   \nn \\
&&
=
\frac{1}{2}
\Big(
\int_q \frac{    \,(p_0 q_0)   }{  q^2 (q+k)^2  (q+k+p)^2  }   
-
\int_q \frac{  p^2  \, (p_0 q_0)  }{  q^2 (q+k)^2  (q+k+p)^2 (q+p)^2 }  \cr
&& -
\int_q \frac{     \, (p_0 q_0)   }{   (q+k)^2  (q+k+p)^2 (q+p)^2 }
\Big)  \cr
&&
=
\frac{1}{2}
\Big(
L(\vec{p}_0; k,k+p)
-
\int_q \frac{  p^2  \, (p \cdot q - \bm{p \cdot q})  }{  q^2 (q+k)^2  (q+k+p)^2 (q+p)^2 } 
-L(\vec{p}_0; k,k-p )
+ p_0^2 \, J_3 (k,k-p)
\Big)   \cr
&&
=
\frac{1}{2}
\Big(
L(\vec{p}_0; k,k+p)
-
\big[
\frac{p^2 }{2} J_3(k,k+p)
- \frac{p^4 }{2} \, J_4(k,p)
- \frac{p^2 }{2}  J_3(k,k-p)
-p^2 \, M(p;k)
\big]    \nn   \\
&&  \qquad 
-L(\vec{p}_0; k,k-p )
+ p_0^2 \, J_3 (k,k-p)
\Big) .
\end{eqnarray}
Note that $M(p;k)$ is not symmetric with respect to the exchange of its arguments.

Next, we consider
\begin{eqnarray}
&&  
 D^{ \rm sym}_{\alpha \beta } (k) 
\frac{ k_\alpha q_\beta [\ldots]+q_\alpha k_\beta [\ldots]}{ q^2 (q+k)^2  (q+k+p)^2 (q+p)^2} 
=
\frac{ 8 (k \cdot q) \big[ f_B (\bm{k \cdot q}) 
+ f_C \, (\bm{k^2})  (\bm{k \cdot q}) +   f_A  \,k_0 \,q_0  \big]  }
{q^2 (q+k)^2  (q+k+p)^2 (q+p)^2 }  \nn \\
&&
+
\frac{   (\bm{k \cdot q})  \big[4 f_B (k \cdot p) + 4 f_C \,(\bm{k^2}) (k \cdot p) 
    + 2 f_B \,p^2 + 2 f_C (\bm{k^2}) \,p^2  \big] }
{q^2 (q+k)^2  (q+k+p)^2 (q+p)^2 } \cr
&&
+
\frac{ [  4 f_A (k \cdot p)  + 2 f_A \,p^2 ] \,  k_0 q_0 }
{q^2 (q+k)^2  (q+k+p)^2 (q+p)^2 } .
\end{eqnarray}
Integrating over $q$ gives
\begin{eqnarray}
&&
[8 f_B + 8 f_C \, \bm{k^2} ] \,F_4 + 8 f_A \, [ M_2(k;k,p) - F_4]   \nn \\
&&
+
 \big[4 f_B (k \cdot p) + 4 f_C \,(\bm{k^2}) (k \cdot p) 
    + 2 f_B \,p^2 + 2 f_C (\bm{k^2}) \,p^2  \big]  \, M(k;p)    \nn    \\
&&
+[  4 f_A (k \cdot p)  + 2 f_A \,p^2 ] \,
 [  \frac{1}{2} J_3(k+p,p)  - \frac{k^2}{2}   J_4(k,p) 
  - \frac{1}{2} J_3 (k,k-p)   -M(k,p) ],
\end{eqnarray}
where, using \eqref{kq2D-kq3D},
\begin{eqnarray}
F_4  \, && \equiv 
\int_q  
\frac{ (k \cdot q)   (\bm{k \cdot q})  }{q^2 (q+k)^2  (q+k+p)^2 (q+p)^2 } \cr
&&
=
\frac{1}{2}
\int_q 
\frac{  [ (q+k)^2 - k^2 -q^2  ] (\bm{k \cdot q})  }{q^2 (q+k)^2  (q+k+p)^2 (q+p)^2 }  \cr
&&
=
 \frac{1}{2}
\Big(
L(\bm{k} ,p,k+p)
- k^2 \,  M(k ; p)
-
L(\bm{k} , p, p-k)
+ \bm{k}^2  \frac{1}{ 8 |p| \, |p-k| \, |  k|   }
\Big).
\end{eqnarray}

We also note the integral relation,
\begin{eqnarray}
F_5 \, && \equiv 
\int_q  
\frac{ (k \cdot q) \,k_0 \,q_0  \big]  }{q^2 (q+k)^2  (q+k+p)^2 (q+p)^2 } \cr
&& =
\int_q  
\frac{ (k \cdot q) \,(k \cdot q- \bm{k \cdot q   })     }{q^2 (q+k)^2  (q+k+p)^2 (q+p)^2 } \cr
&& =
M_2(k;k,p) - F_4.
\qquad  \qquad 
\end{eqnarray}
where the definition in Eq.~\eqref{M2-Fun} for $M_2$ was used.
In fact, $M_2(k;k,p)$ can be simplified further in terms of $J_2,J_3, J_4, I$.

Finally, we consider
\begin{eqnarray}
&&
 D^{ \rm sym}_{\alpha \beta } (k) 
\frac{ + p_\alpha k_\beta [\ldots]+   k_\alpha p_\beta [\ldots] }{q^2 (q+k)^2  (q+k+p)^2 (q+p)^2}
=
\frac{ (k \cdot q) \; 
[4 f_B (\bm{k \cdot p}) + 4 f_C \bm{k}^2  (\bm{k \cdot p}) + 4 f_A\, k_0 p_0 ]  } 
{q^2 (q+k)^2  (q+k+p)^2 (q+p)^2} \cr
&& +
\frac{(p \cdot q) \; [-2 f_B  (\bm{k \cdot p}) - 2 f_C \bm{k}^2  (\bm{k \cdot p}) 
                                 - 2 f_A  \,k_0 p_0  ] }
{q^2 (q+k)^2  (q+k+p)^2 (q+p)^2} .
\end{eqnarray}
Integrating over $q$ gives
\begin{eqnarray}
&&
[4 f_B (\bm{k \cdot p}) + 4 f_C \bm{k}^2  (\bm{k \cdot p}) + 4 f_A\, k_0 p_0 ] 
\,\times
\frac{1}{2}[ J_3(k+p,p)- k^2 J_4(p,k) - J_3 (k,k-p)  ]   \nn \\
&& 
+
 [-2 f_B  (\bm{k \cdot p}) - 2 f_C \bm{k}^2  (\bm{k \cdot p}) - 2 f_A  \,k_0 p_0  ]
\times 
\frac{1}{2} [J_3(k+p,k) - p^2 J_4(p,k) - \cr
&&J_3(k,k-p)   ]                                 
\end{eqnarray}

Putting these results together produces $I_a$ (schematically):
\begin{eqnarray}
I_{a}
\equiv
\int \frac{d^3q}{(2\pi)^3}
\Tr[...   ]_{\mu \nu}  (  \frac{ \delta_{\mu \nu} }{ 2  } ) D^{\text{Sym}}_{\alpha \beta } (k) .
\label{Pi-a-Sym}
\end{eqnarray}
The result is a lengthy expression that we do not write out here.

\subsection{Combining the symmetric components of $\Pi^{(a)}$ and $\Pi^{(b+c)}$} 

We add together \eqref{Pi-a-Sym} and \eqref{Pi-bc-Sym} to find (including multiplying \eqref{Pi-bc-Sym} by a factor of 2) the symmetric component:
\begin{eqnarray}
&&   
\Pi^{(a+b+c)}_{(\text{S}) }(p)= 
\frac{1}{2} \delta_{\mu \nu}  \Pi^{(a+b+c)}_{\mu \nu} (p)   
=
\frac{(-1) g^4}{N_f}  
\int \frac{d^3k}{(2\pi)^3}   ( 2 \, I_{bc} + I_{a} )    \nn \\
&& 
=
\frac{(-1) g^4}{N_f}
\int_{-1}^{1} d\cos \theta  
\int_0^{\Lambda} dk \frac{2\pi k^2}{(2\pi)^3}   ( 2 \, I_{bc} + I_{a} )  .
\end{eqnarray}
As before, we take the external momentum $p$ to lie along the $k_\tau$ direction so that $k \cdot p = |k| |p| \cos \theta$. 
This choice allows us to set all $\bm{k \cdot p} =0$ in $2 \, I_{bc} + I_{a}$ expression.
We plug in the propagator in \eqref{DSym-parameters}, perform the $dk $ radial direction integral with cut off $\Lambda$, and perform a 
$\frac{1}{\Lambda}$ expansion to find:
\begin{eqnarray}
\Pi^{(a+b+c)}_{(\text{S}) }(p) 
=
- \frac{g^4}{N_f}
|p|
\int_{-1}^{1} d\cos \theta  
\Big(
f^{\wx}_0 (\cos\theta)
+
f^{\wx}_{L} (\cos\theta) \, \log[\frac{\Lambda}{|p|}]
+
f^{\wx}_{1} (\cos\theta) \, \Lambda
\Big). \quad \quad
\label{f0-fL-f1-def}
\end{eqnarray}
$f^{\wx}_{L}, ~ f^{\wx}_1$ are odd functions of $\cos \theta$ so they both vanish after performing the angular integration.
Symmetrizing $f^{\wx}_0(\cos\theta)$ to remove any antisymmetric part, we find
\begin{eqnarray}
&& 
\frac{f^{\wx}_0(z) + f^{\wx}_0(-z)  }{2}      \\
&&   
\equiv 
\frac{
\Big(
-5 \,C_Y \,  z (-1 + z^2) + z \sqrt{1-z^2} (9 + z^2) + 
   6 (-1 + z^2) (C_Y - C_Y\,  z^2 + 2 \sqrt{1-z^2}) \text{arctanh}[z]
\Big)
}
{ 192 \pi^2 \, z \, [\; A_Y\sqrt{1-z^2} + B_Y(1-z^2)    \;\;  ]}   , \nn
\end{eqnarray}
where $z = \cos \theta$.
Thus, we have
\begin{eqnarray}
&&
\sigma^{\text{2loop-gauge}}_{xx}  
 =
\frac{i}{\omega}   \Pi^{(a+b+c)}_{(\text{S}) }  (\bm{p} = 0, |p_0| \to i \omega) 
 =
 \frac{(-1) g^4}{N_f}
\int_{-1}^{1} dz
\frac{f^{\wx}_0(z) + f^{\wx}_0(-z)  }{2  }      .
\end{eqnarray}
This agrees with Eq.~\eqref{fwx-avg} in the main text.

As a consistency check, we note that our result agrees with \cite{Huh:2015aa, Giombi:2016fct} when $\wx=0$:
\begin{eqnarray}
\frac{f^{\wx=0}_0(z) + f^{\wx=0}_0(-z)  }{2}
=
\frac{ \gX^2 }{\kappa^2+ \gX^4     }   \frac{ 1  }{192 \pi^2 \, z}
\Big(
9z + z^3+ 12(z^2 -1) \text{arctanh}[z]
\Big) 
\label{f0-expicit-form}
\end{eqnarray}
and so
\begin{eqnarray}
\int_{-1}^{1} d\cos \theta  
f^{\wx=0}_0 (\cos\theta) \Big{|}_{\kappa =0, g^2_X = \frac{1}{16}, g =1  }
\approx
0.00893191    
=
\frac{1}{16} \times 0.14291062 .
\end{eqnarray}

\bibliography{tmpMybib}


\end{document}